\documentclass[pre,twocolumn,superscriptaddress]{revtex4}
\usepackage{graphicx,epsfig}

\DeclareGraphicsExtensions{. jpg,. pdf, . mps, .png, .eps, . ps, . EPS}

\usepackage{amsmath}
\usepackage{color}
\begin{document}
\newcommand{\beq}{\begin{equation}}
\newcommand{\eeq}{\end{equation}}
\newcommand{\bea}{\begin{eqnarray}}
\newcommand{\eea}{\end{eqnarray}}
\newcommand{\gt}{\tilde{g}}
\newcommand{\mt}{\tilde{\mu}}
\newcommand{\et}{\tilde{\varepsilon}}
\newcommand{\ct}{\tilde{C}}
\newcommand{\bt}{\tilde{\beta}}

\newcommand{\avg}[1]{\langle{#1}\rangle}
\newcommand{\Avg}[1]{\left\langle{#1}\right\rangle}
\newcommand{\cor}[1]{\textcolor{red}{#1}}

\title{Weighted Multiplex Networks}
\author{Giulia Menichetti}
\affiliation{Department of Physics and Astronomy, Bologna University, Viale B. Pichat 6/2 40127 Bologna, Italy}
\author{Daniel Remondini}
\affiliation{Department of Physics and Astronomy, Bologna University, Viale B. Pichat 6/2 40127 Bologna, Italy}
\author{Pietro Panzarasa}
\affiliation{School of Business and Management, Queen Mary University of London, London E1 4NS , United Kingdom}
\author{Ra\'ul J. Mondrag\'on}
\affiliation{School of Electronic Engineering and Computer Science, Queen Mary University of London, London E1 4NS, United Kingdom}
\author{Ginestra Bianconi}
\affiliation{School of Mathematical Sciences, Queen Mary University of London, London E1 4NS, United Kingdom}
\begin{abstract}
One of the most important challenges in network science is to quantify the information encoded in complex network structures. Disentangling randomness from organizational principles is even more demanding when networks have a multiplex nature. Multiplex networks are multilayer systems of $N$ nodes that can be linked in multiple interacting and co-evolving layers. In these networks, relevant information might not be captured if the single layers were analyzed separately. Here we demonstrate that such partial analysis of layers fails to capture significant correlations between weights and topology of complex multiplex networks. To this end, we study two weighted multiplex co-authorship and citation networks involving the authors included in the American Physical Society. We show that in these networks weights are strongly correlated with multiplex structure, and provide empirical evidence in favor of the advantage of studying weighted measures of multiplex networks, such as multistrength and the inverse multiparticipation ratio. Finally, we introduce a theoretical framework based on the entropy of multiplex ensembles to quantify the information stored in multiplex networks that would remain undetected if the single layers were analyzed in isolation.
\end{abstract}
\maketitle

{\bf A large variety of systems, including social, infrastructure, and biological ones, can be described as multiplex networks in which pairs of nodes can be connected through multiple links across multiple layers. Over the past fifteen years scientists have investigated  the single layers of these networks, thus neglecting to uncover the information encoded in their multiplex nature. Here we focus on weighted multiplex networks, and provide evidence that novel information can be extracted when the interacting and co-evolving layers are taken into account. To this end, we propose a new indicator based on the entropy of multiplex ensembles for quantifying the amount of information that would remain undetected if single layers of multiplex networks were analyzed in isolation.}\\

{N}etwork theory  investigates the global topology and organization structure of graphs formed by individual interactions among the constituent elements of a number of complex systems including social groups, infrastructure and technological systems, the brain and biological networks \cite{RMP, Newmanrev, Boccaletti2006, Fortunato}. Over the last fifteen years, a large body of literature has attempted to disentangle noise and stochasticity from non-random patterns and mechanisms, in an attempt to gain a better understanding of how these systems function and evolve. More recently, further advances in the study of complex systems have been spurred by the upsurge of interest in multiplex networks in which pairs of interacting elements are represented as nodes connected through multiple types of links, at multiple points in time, or at multiple scales of resolution \cite{Arenas_rev}. More specifically, a multiplex network is a set of $N$ nodes interacting in $M$ layers, each reflecting a distinct type (or time or resolution) of interaction linking the same pair of nodes. Examples of multiplex networks include: social networks, where the same individuals can be connected through different types of social ties originating from friendship, collaboration, or family relationships \cite{Thurner}; air transportation networks, where different airports can be connected through flights of different companies \cite{Boccaletti}; and the brain, where different regions can be seen as connected by the functional and structural neural networks \cite{Bullmore}.

Most of the studies so far conducted on multiplex networks have been concerned with the empirical analysis of a wide range of systems \cite{Thurner, Boccaletti, Kurths, Barthelemy}, with modeling their underlying structures \cite{Latora, MPageRank, Mucha}, and with describing new critical phenomena and processes occurring on them \cite{Havlin1, Dorogovtsev, Diffusion, Leicht}. Despite the growing interest in multiplex networks, a fundamental question still remains largely unanswered: What is the advantage of a full-fledged analysis of complex systems that takes all their interacting layers into account, over more traditional studies that represent such systems as single networks with only one layer? To answer this question, one should demonstrate that novel and relevant information can be uncovered only by taking the multiplex nature of complex systems directly into account, and would instead remain undetected if individual layers were analyzed in isolation. In this paper, an attempt is made to offer a possible solution to this problem within the context of weighted multiplex networks.

Like with single networks, links between nodes may have a different weight, reflecting their intensity, capacity, duration, intimacy or exchange of services \cite{Granovetter}. The role played by the weights in the functioning of many networks, and especially the relative benefits of weak and strong ties in social networks, have been the subject of a longstanding debate \cite{Granovetter, Barabasiweights, Vespignaniweights}. Moreover, it has been shown that, in single networks, the weights can be distributed in a heterogeneous way, as a result of the non-trivial effects that the structural properties of the networks have on them \cite{Barrat}. In particular, correlations between weights and structural properties of single networks can be uncovered by the analysis of strength-degree correlations and by the distribution of the weights of the links incident upon the same node. To characterize weighted networks, it is common practice to measure the following quantities: i) the average strength of nodes of degree $k$, i.e. $s=s(k)$, describing how weights are distributed in the network; and ii) the average inverse participation ratio of the weights of the links incident upon nodes of degree $k$, i.e. $Y=Y(k)$, describing how weights are distributed across the links incident upon nodes of degree $k$. Here we show that these two quantities do not capture the full breadth of the information encoded in multiplex networks. Indeed, a full-fledged analysis of the properties of multiplex networks is needed that takes the multiple interacting and co-evolving layers simultaneously into account. 

\begin{figure}[t]
\begin{center}
\centerline{\includegraphics[width=3.3in]{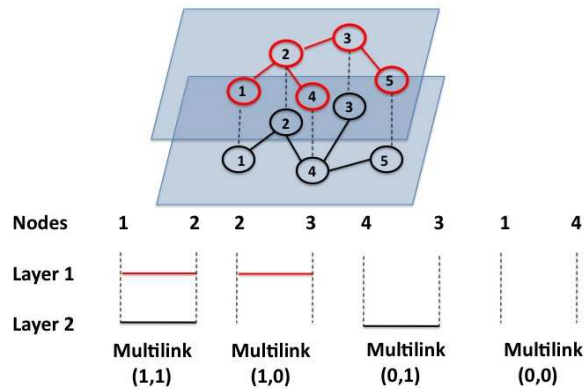}}
 \end{center}
  \caption{\label{fig:fig1} Example of all possible multilinks in a multiplex network with $M=2$ layers and $N=5$ nodes. Nodes $i$ and $j$ are linked by one multilink $\vec{m}=(m_{\alpha},m_{\alpha '})$. 
}
\end{figure}
For a multiplex network, a {\em multilink} $\vec{m}=(m_1,m_2,\ldots, m_M)$ between nodes $i$ and $j$ indicates the set of all links connecting these nodes in the different layers \cite{BianconiPRE}. In particular, if $m_{\alpha}=1$, there is a link between nodes $i$ and $j$ in layer $\alpha$, whereas if $m_{\alpha}=0$ nodes $i$ and $j$ are not connected in layer $\alpha$. Multilink $\vec{m}=\vec{0}$ between two nodes refers to the case in which no link exists between the two nodes in all layers of the multiplex network. Thus, multilinks indicate the most straightforward type of correlation between layers, and provide a simple generalization of the notion of overlap. In fact, if nodes $i$ and $j$ are connected by a multilink $\vec{m}$ , with $m_{\alpha}=m_{\alpha'}=1$, it follows that there is an overlap of links between $i$ and $j$ in layers $\alpha$ and $\alpha'$. Figure~\ref{fig:fig1} shows a multiplex network with $M=2$ layers and $N=5$ nodes with different types of multilinks.

Here we will define two new measures, {\em multistrength} and the {\em inverse multiparticipation ratio}, which are, respectively, the sum of the weights of a certain type of multilink incident upon a single node and a way for characterizing the heterogeneity of the weights of multilink $\vec{m}$ incident upon a single node. To provide empirical evidence that weighted properties of multilinks are fundamental for properly assessing weighted multiplex networks, we focus on the networks of the authors of papers published in the journals of the American Physical Society (APS), and analyze the scientific collaboration network and the citation network connecting the same authors. These networks are intrinsically weighted since any two scientists can co-author more than one paper and can cite each other's work several times. A large number of studies have analyzed similar bibliometric datasets drawing upon network theory \cite{Redner, Newman,Newman2001, Santo_APS, Santo_citations}. Unlike these studies, here we investigate the APS bibliometric dataset using the framework of multiplex networks that allows us to explore novel properties of the collaboration and citation networks. In particular, we show that multistrength and the inverse multiparticipation ratio enable new relevant information to be extracted from the APS dataset and that this information extends beyond what is encoded in the strength and inverse participation ratio of single layers. Finally, based on the entropy of multiplex ensembles, we propose an indicator $\Xi$ to evaluate the additional amount of information that can be extracted from the weighted properties of multilinks in multiplex networks over the information encoded in the properties of their individual layers analyzed separately. 

\section{Weighted multiplex networks}
\subsection{Definition} 
A weighted multiplex network is a set of $M$ weighted networks $G_{\alpha}=(V,E_{\alpha})$, with $\alpha=1,\ldots,M$. The set of nodes $V$ is the same for each layer and has cardinality $|V|=N$, whereas the set of links $E_{\alpha}$ depends on the layer $\alpha$. A multiplex network is represented formally as $\vec{G}=(G_1,G_2,\ldots, G_{\alpha},\ldots G_M)$.
Each network $G_{\alpha}$ is fully described by the adjacency matrix 
${\bf a}^\alpha$ with elements $a^{\alpha}_{ij}$, where $a_{ij}^{\alpha}=w_{ij}^{\alpha}>0$ if there is a link with weight $w_{ij}^{\alpha}$ between nodes $i$ and $j$ in layer $\alpha$, and 
$a_{ij}^{\alpha}=0$ otherwise. From now on, in order to simplify the formalization of weighted multiplex networks, we will assume that the weight of the link between any pair of nodes $i$ and $j$, 
$a_{ij}^{\alpha}=w_{ij}^{\alpha}$, can only take integer values. This does not represent a major limitation because in a large number of weighted multiplex networks the weights of the links can be seen as multiples of a minimal weight.
\subsection{Structural properties of individual layers}
We indicate the degree of node $i$ in layer $\alpha$ with 
$k_i^{\alpha}$, 
defined as 
$k_i^{\alpha}=\sum_{j=1}^N \theta(a_{ij}^{\alpha}),$
where function $\theta(x)=1$ if $x>0$ and $\theta(x)=0$ otherwise. In complex weighted networks, weights can be distributed across links more or less heterogeneously. A way to evaluate this heterogeneity is to introduce local properties such as the {\em strength} $s_{i}^\alpha$ and the {\em inverse participation ratio} $Y_i^{\alpha}$ of node $i$ in layer $\alpha$: 
\begin{eqnarray}
s_i^{\alpha}&=&\sum_{j=1}^N a_{ij}^{\alpha}, \nonumber \\
Y_i^{\alpha}&=&\sum_{j=1}^N \left(\frac{a_{ij}^{\alpha}}{s_i^{\alpha}}\right)^2.
\end{eqnarray}
Like with single networks, in any given layer $\alpha$, the strength 
$s_i^{\alpha}$ of a node indicates the sum of the weights of the links incident upon node $i$ in layer $\alpha$, whereas the inverse participation ratio $Y_i^{\alpha}$ indicates how unevenly the weights of the links of node $i$ are distributed in layer $\alpha$. The inverse of $Y_i^{\alpha}$ characterizes the effective number of links of node $i$ in layer $\alpha$. Indeed, $(Y_i^{\alpha})^{-1}$ is greater than one and smaller than the degree of node $i$ in layer $\alpha$, i.e. $(Y_i^{\alpha})^{-1}\in(1,k_i^{\alpha})$. Moreover, if the weights of the links of node $i$ are distributed uniformly, i.e. $w_{ij}^{\alpha}=s_i^{\alpha}/k_i^{\alpha}$, we have $(Y_i^{\alpha})^{-1}=k_i^{\alpha}$. Conversely, if the weight of one link is much larger than the other weights, i.e. $w_{ir}^{\alpha}\gg w_{ij}^{\alpha}$ for every $j\neq r$, then $(Y_i^{\alpha})^{-1}=1$.

In network theory, it is common practice to evaluate the conditional means of the strength and of the inverse participation ratio of the weights of links against the degree of nodes. In a multiplex network, we will then consider the quantities 
$s^{\alpha}(k)=\Avg{ s_i^{\alpha} \delta(k_i^{\alpha},k)}$ and $Y^{\alpha}(k)=\Avg{ Y_i^{\alpha} \delta(k_i^{\alpha},k)}$, where the average is calculated over all nodes with degree $k$ in layer $\alpha$, and $\delta(a,b)$ indicates the Kronecker delta. Like in single networks \cite{Barrat}, $s_k^{\alpha}$ is expected to scale as 
\begin{eqnarray} s^{\alpha}(k)\propto k^{\beta_{\alpha}},
\label{Sksingle}
\end{eqnarray} 
with $\beta_{\alpha}\ge1$. We can distinguish between two scenarios. In the first one, the average strength of nodes with degree $k$ increases linearly with $k$, i.e. $\beta_{\alpha}=1$. This indicates that, on average, the weights of the links incident upon the hubs do not differ from the weights of the links of less connected nodes. In the second scenario, the strength of the nodes with degree $k$ increases super-linearly with $k$, i.e. $\beta_{\alpha}>1$, thus indicating that, on average, the weights of the links incident upon the hubs are larger than the weights of the links of less connected nodes. In a multiplex network, it may be the case that weights are distributed in different ways across the layers. For instance, some layers may be characterized by a super-linear growth of $s_k^{\alpha}$, while other layers may show a linear dependence. Finally, the inverse participation ratio can be used in order to characterize the heterogeneity of the weights of the links incident upon nodes with a certain degree. In particular, it has been observed that, in many single weighted networks, the inverse participation ratio scales as an inverse power-law function of the degree of nodes. In a multiplex network, this would imply 
\begin{eqnarray}
Y^{\alpha}(k)\propto \frac{1}{k^{\lambda_{\alpha}}},
\label{Yksingle}
\end{eqnarray}
where exponent $\lambda_{\alpha}\leq 1$ is layer-dependent.

\subsection{Multilink, multistrength, and inverse multiparticipation ratio}
A number of multiplex networks are characterized by a significant overlap of links across the different layers~\cite{Thurner, Boccaletti}. In order to generalize the notion of overlap to weighted multiplex networks, in what follows we will draw on the concept of multilink~\cite{BianconiPRE}. Let us consider the vector $\vec{m}=(m_1, m_2,\ldots, m_{\alpha},\ldots, m_M)$ in which every element $m_{\alpha}$ can take only two values $m_{\alpha}=0,1$. We define a {\em multilink} $\vec{m}$ the set of  links connecting a given pair of nodes in the different layers of the multiplex and connecting them in the generic layer $\alpha$ only if $m_{\alpha}=1$. In particular two nodes $i$ and $j$ are always linked by a single multilink of type $\vec{m}=\vec{m}^{ij}=(\theta(a_{ij}^1),\theta(a_{ij}^2),\ldots,\theta(a_{ij}^M) )$, 
where $\theta(x)=1$ if $x>0$, and $\theta(x)=0$ otherwise. The multilink $\vec{m}=\vec{0}$ between two nodes represents the situation in which in all the layers of the multiplex the two nodes are not directly linked.
We can therefore introduce the  multiadjacency matrices ${\bf A}^{\vec{m}}$ with elements $A^{\vec{m}}_{ij}$ equal to 1 if there is a multilink $\vec{m}$ between node $i$ and node $j$ and zero otherwise.\\
In terms of the weighted adjacency matrices ${\bf a}^{\alpha}$ of the multiplex
 the elements $A^{\vec{m}}_{ij}$ of the multiadjacency matrix ${\bf A}^{\vec{m}}$ are given by 
\begin{equation}
A^{\vec{m}}_{ij}=\prod_{\alpha=1}^{M}[\theta(a_{ij}^{\alpha})m_{\alpha}+ (1-\theta(a_{ij}^{\alpha}))(1-m_{\alpha})]
\label{multilink}
\end{equation}
where $\theta(x)=1$ if $x>0$, otherwise $\theta(x)=0$.
Even though there are $2^M$  multiadjacency matrices, only $2^M-1$ of them are independent because the normalization condition, $\sum_{\vec{m}}A_{ij}^{\vec{m}}=1$, must be satisfied for any pair of nodes $i$ and $j$. Based on multi-adjacency matrices, we can define the {\em multidegree} 
$k_i^{\vec{m}}$ of node $i$ as 
\begin{eqnarray}
k_i^{\vec{m}}=\sum_{j=1}^N A_{ij}^{\vec{m}},
\end{eqnarray}
which indicates how many multilinks $\vec{m}$ are incident upon node $i$.

To study weighted multiplex networks, we now introduce two new measures. For layer $\alpha$ associated to multilinks $\vec{m}$, such that $m_{\alpha}>0$, we define the multistrength $s^{\vec{m}}_{i,\alpha}$ and the inverse multiparticipation ratio $Y^{\vec{m}}_{i,\alpha}$ 
of node $i$, respectively, as 
\begin{eqnarray}
s^{\vec{m}}_{i,\alpha}=\sum_{j=1}^N a_{ij}^{\alpha}A_{ij}^{\vec{m}},
\end{eqnarray}
\begin{eqnarray}
Y^{\vec{m}}_{i,\alpha}=\sum_{j=1}^N \left(\frac{a_{ij}^{\alpha}A_{ij}^{\vec{m}}}{\sum_r a_{ir}^{\alpha} A_{ir}^{\vec{m}}}\right)^2.
\end{eqnarray}
Since there are $\binom{M}{k}$ multilinks $\vec{m}$ such that $\sum_{\alpha}m_{\alpha}=k$, for each node the number of multistrengths that can be defined in a multiplex network of $M$ layers is $K=M2^{M-1}$. The average multistrength of nodes with a given multidegree, i.e $s^{\vec{m},\alpha}(k^{\vec{m}})=\Avg{s_i^{\vec{m},\alpha}\delta(k_i^{\vec{m}},k^{\vec{m}})}$, and the average inverse multiparticipation ratio of nodes with a given multidegree, $Y^{\vec{m},\alpha}(k^{\vec{m}})=\Avg{Y_i^{\vec{m},\alpha}\delta(k_i^{\vec{m}},k^{\vec{m}})}$, are expected to scale as  
\bea
s^{\vec{m},\alpha}(k^{\vec{m}}) &\propto &(k^{\vec{m}})^{\beta_{\vec{m},\alpha}} \nonumber \\
Y^{\vec{m},\alpha}(k^{\vec{m}})&\propto & \frac{1}{(k^{\vec{m}})^{\lambda_{\vec{m},\alpha}}},
\label{SkYk}
\eea
with exponents $\beta_{\vec{m},\alpha} \geq 1$ and $\lambda_{\vec{m},\alpha}\leq 1$.
The use of multilinks $\vec{m}$ to describe multiplex properties is numerically feasible if the number of layers is smaller than the number of nodes, i.e. $M\ll \log(N)$. If this condition is not satisfied, then the following quantities can be measured: the {\em overlap multiplicity}, $\nu(\vec{m})=\sum_{\alpha}m_{\alpha}$, which indicates that multilink $\vec{m}$ connects two nodes through $\nu(\vec{m})$ links; $s^{\alpha}(\nu)=\Avg{s_{i,\alpha}^{\vec{m}}}_{\nu({\vec{m}})=\nu}$; and $Y^{\alpha}(\nu)=\Avg{Y_{i,\alpha}^{\vec{m}}}_{\nu({\vec{m}})=\nu}$, where $\nu=1,2\ldots, M$.

\section{Empirical evidence of weighted properties of multilinks}
In this section, we will draw on the measures introduced above and provide empirical evidence that, in weighted multiplex networks, weights can be correlated with the multiplex structure in a non-trivial way. To this end, we analyzed the bibliographic dataset that includes all articles published in the APS journals (i.e., { Physical Review Letters}, { Physical Review}, and { Reviews of Modern Physics}) from 1893 to 2009. Of these articles, the dataset includes their citations as well as the authors. Here, we restrict our study only to articles published either in {Physical Review Letters (PRL) or in Physical Review E (PRE) and written by ten or fewer authors, $n_p\leq 10$. We constructed multiplex networks in which the nodes are the authors and links between them have a two-fold nature: scientific collaborations with weights defined as in \cite{Newman2001} (see Supplementary Material for details), and citations with weights indicating how many times author $i$ cited author $j$.

In particular, we created the following two duplex networks (i.e., multiplex networks with $M=2$):
\begin{enumerate}
 \item
{\bf  CoCo-PRL/PRE: }{\em collaborations among PRL and PRE authors.}
The nodes of this multiplex network are the authors with articles published both in {PRL} and { PRE} (i.e., $16,207$ authors). These nodes are connected in layer $1$ through weighted undirected links indicating the strength of their collaboration in { PRL} (i.e., co-authorship of PRL articles). The same nodes are connected in layer $2$ through weighted undirected links indicating the strength of their collaboration in { PRE} (i.e., co-authorship of { PRE} articles).
\item
{\bf CoCi-PRE: }{\em collaborations among { PRE} authors and citations to { PRE} articles.} The nodes of this multiplex network are the authors of articles published in { PRE} (i.e., $35,205$ authors). These nodes are connected in layer $1$ through weighted undirected links indicating the strength of their collaboration in { PRE} (i.e., co-authorship of { PRE} articles). The same nodes are connected in layer $2$ through weighted directed links indicating how many times an author (with articles in { PRE}) cited another author's work, where citations are limited to those made to { PRE} articles.
\end{enumerate}

Both these multiplex networks show a significant overlap of links and a significant correlation between degrees of nodes as captured by the Pearson correlation coefficient $\rho$ (see Supplementary Material). This finding supports the hypothesis that the two layers in each of the multiplex networks are correlated. That is, the existence of a link between two authors in one layer is correlated with the existence of a link between the same authors in the other layer. Moreover, the multidegrees of the multiplex networks are broadly distributed, and the hubs in the scientific collaboration network tend to be also the hubs in the citation network (see Supplementary Material).

In the case of the CoCo--PRL/PRE network, multilinks $\vec{m}=(1,0)$, $\vec{m}=(0,1)$ and $\vec{m}=(1,1)$ refer to collaborations only in { PRL}, only in { PRE}, and in both { PRL} and { PRE}, respectively. Moreover, to distinguish the weights used when evaluating multistrength, we have $\alpha=PRL$ or $\alpha=PRE$. Results indicate that the multistrengths and inverse multiparticipation ratio behave according to Eq. $(\ref{SkYk})$ (see Fig.~\ref{fig2}). The difference between exponents $\beta_{\vec{m},PRL}$ for $\vec{m}=(1,0)$ and $\vec{m}=(1,1) $ is not  statistically significant. Nevertheless, there is a statistically significant difference between the average weights of multilinks $(1,0)$ and $(1,1)$ in the { PRL} layer. As to the inverse multiparticipation ratio, there is a significant variation in the exponents, $\lambda_{(1,0),PRL}=0.84 \pm 0.03$ and  $\lambda_{(1,1),PRL}=0.74 \pm 0.05$ (see Fig.~\ref{fig2}, bottom left panel). This suggests that the weights of the collaborative links between co-authors of both { PRL} and { PRE} articles are distributed more heterogeneously than the weights of collaborative links between co-authors of articles published only in { PRL} (see Supplementary Material for details on our statistical tests). Similar results were found for multistrengths evaluated in the { PRE} layer (see Fig.~\ref{fig2}, right panels).  

These findings clearly indicate that the partial analysis of individual layers would fail to uncover the fact that the average weight of the link between authors that collaborated both on { PRL} and { PRE} articles is significantly larger than the average weight of the link between authors that collaborated only on articles published in one journal. Moreover, the difference in functional behavior of the multipartition ratio across layers could not be captured if layers were analyzed separately.

\begin{figure}[t]
\begin{center}
 \centerline{\includegraphics[width=3.0in]{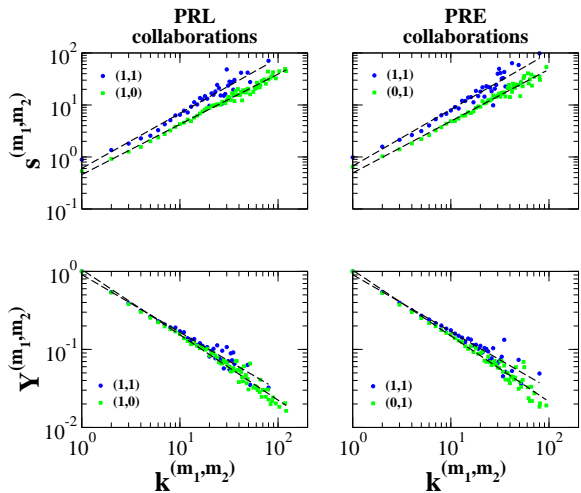}}
 \end{center}
\caption{\label{fig2}Average multistrength and average inverse multiparticipation ratio versus multidegree in the CoCo-PRE/PRL multiplex network. The average multistrengths and the average inverse multiparticipation ratio are fitted by a power-law distribution of the type described in Eq. $(\ref{SkYk})$ (fitted distributions are here indicated by black dashed lines). Statistical tests for the collaboration network of PRL suggest that the exponents $\beta_{\vec{m},1}$ defined in Eq. $(\ref{SkYk})$ are the same, while exponents $\lambda_{\vec{m},PRL}$ are significantly different. Similar results can be obtained for the exponents in the PRE collaboration layer. Nevertheless, multistrengths $s^{(1,1),\alpha}$ are always larger that multistrengths $s^{(1,0),PRL}$ and $s^{(0,1),PRE}$, when multistrengths are calculated over the same number of multilinks, i.e. $k^{(1,1)}=k^{(1,0)}=k^{(0,1)}$ (see Supplementary Material for the statistical test on this hypothesis).}
\end{figure}

In the case of the CoCi-PRE network there are even more significant differences between the properties of the multilinks than in the previous network. In the CoCi-PRE network the functional behavior of multistrength also depends on the type of multilink. Figure~\ref{fig3} shows the average multistrength in the CoCi-PRE network. To distinguish between the weights used to measure multistrength, we have layer $\alpha=col$, which refers to the collaboration network constructed on { PRE} articles, and layer $\alpha=cit$, which refers to the citation network between PRE articles, where a distinction is also made between incoming ($in$) and outgoing ($out$) links. First, in the scientific collaboration network, exponents $\beta_{\vec{m},col}$ are not statistically different, but the average weight of multilink $(1,1)$ is larger than the average weight of multilinks $(1,0),in$ and $(1,0)out$. Moreover, exponents $\lambda_{(1,0),col,in}$ and $\lambda_{(1,0),col,out}$ are larger than exponents $\lambda_{(1,1),col,in},\lambda_{(1,1),col,out}$, indicating that the weights of authors' collaborative links with other cited/citing authors are distributed more heterogeneously than the weights of authors' collaborative links with other authors with whom there are no links in the citation network. Second, in the citation network multistrengths follow a distinct functional behavior depending on the different type of multilink, and are characterized by different $\beta_{\vec{m},cit,in/out}$ exponents. In fact the fitted values of these exponents are given by $\beta_{(1,1)cit,,in}=1.30 \pm 0.07,\beta_{(1,1),cit,out}=1.32 \pm 0.08,\beta_{(0,1,)cit,in}= 1.11 \pm 0.01,\beta_{(0,1),cit,out}=1.10 \pm 0.02$. This implies that, on average, highly cited authors are cited by their co-authors to a much greater extent than is the case with poorly cited authors. A similar, though much weaker effect was also found for the citations connecting authors that are not collaborators. Furthermore, in the citation layer the inverse multiparticipation ratio for multilink $(1,1)$ is always larger than the inverse multiparticipation ratio for multilinks $(1,0)$ and $(0,1)$ (see Supplementary Material for details on the statistical test). Finally, when single layers were analyzed separately, we found $\beta_{col}=1.03 \pm 0.04$ in the collaboration network, and $\beta_{cit,in}=1.13 \pm 0.02$ and $\beta_{cit,out}=1.14 \pm 0.03$ in the citation network. This indicates that in the citation network strength grows super-linearly as a function of degree, i.e. weights are not distributed uniformly. Nevertheless, correlations between weights and types of multilinks cannot be captured if the two individual layers are studied separately.

\begin{figure}[t]
\begin{center}
 \centerline{\includegraphics[width=3.0in]{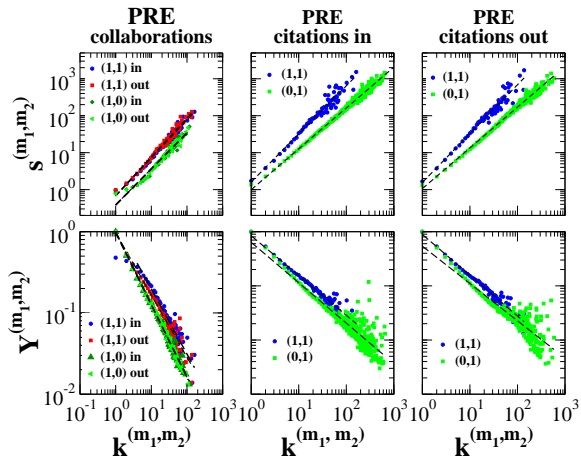}}
 \end{center}
  \caption{\label{fig3} Properties of multilinks in the weighted CoCi-PRE multiplex network. In the case of the collaboration network, the distributions of multistrengths versus multidegrees always have the same exponent, but the average weight of multilinks $(1,1)$ is larger than the average weight of multilinks $(1,0)$. Moreover, the exponents $\lambda_{(1,0),col,in}$, $\lambda_{(1,0),col,out}$ are larger than exponents $\lambda_{(1,1),col,in},\lambda_{(1,1),col,out}$. In the case of the citation layer, both the incoming multistrengths and the outgoing multistrengths have a functional behavior that varies depending on the type of multilink. Conversely, the average inverse multiparticipation ratio in the citation layer does not show any significant change of behavior when compared across different multilinks.}
\end{figure}

\subsection{Assessing the informational content of weighted multilinks}
Recent research on single networks has shown that the entropy of network ensembles provides a very powerful tool for quantifying their complexity \cite{Newman1,Munoz,PNAS}. Here, we propose a theoretical framework based on the entropy of multiplex ensembles for assessing the amount of information encoded in the weighted properties of multilinks. Multiplex weighted network ensembles can be defined as the set of all weighted multiplex networks satisfying a given set of constraints, like for example the expected degree sequence and the expected strength sequence in every layer of the multiplex network, or the expected multidegree sequence and the expected multistrength sequence. A set of constraints imposed upon the multiplex network ensemble uniquely determines the probability $P(\vec{G})$ of the multiplex networks in the ensemble (see Materials and Methods). The entropy ${\cal S}$ of the multiplex ensemble can be defined in terms of $P(\vec{G})$ as 
\bea
{\cal S}=-\sum_{\vec{G}}P(\vec{G})\log P(\vec{G}),
\label{entropy}
\eea 
where ${\cal S}$ indicates the logarithm of the typical number of multiplex networks in the ensemble. The smaller the entropy, the larger the amount of information stored in the constraints imposed on the network. The entropy can be regarded as an unbiased way to evaluate the informational value of these constraints.

In order to gauge the information encoded in a weighted multiplex network with respect to a null model, we define the indicator $\Psi$, which quantifies how much information is carried by the weight distributions of a weighted multiplex ensemble. In particular, $\Psi$ compares the entropy of a weighted multiplex ensemble $\cal S$ with the entropy of a weighted multiplex ensemble in which the weights are distributed homogeneously. Therefore, $\Psi$ can be defined as   
\bea
\Psi=\frac{|{\cal S}-\avg{\cal S}_{\pi(w)}|}{\Avg{(\delta {\cal S})^2}_{\pi(w)}},
\label{psi}
\eea
where $\Avg{(\delta {\cal S})^{2}}_{\pi(w)}$ is the standard deviation, and the average $\avg{\ldots}_{\pi(w)}$ is calculated over multiplex networks with the same structural properties but with weights distributed homogeneously. In particular, when the weight distribution is randomized, the multiplex networks are constrained in such a way that each link must have a minimal weight (i.e. $w_{ij}\ge 1$), while the remaining of the total weight is distributed randomly over the links.

In order to evaluate the amount of information encoded in the weight of links in single layers and compare it to the information supplied by multistrength, we consider the following undirected multiplex ensembles:    
\begin{itemize}
\item {\it  Correlated weighted multiplex ensemble.}
In this ensemble, we fix  the expected multidegree sequence $\{k_i^{\vec{m}}\}$ and we set the expected multistrength sequence $\{s_i^{\vec{m},\alpha}\}$ to be
\bea
s_i^{\vec{m},\alpha}=c_{\vec{m},\alpha}(k^{\vec{m},\alpha})^{\lambda_{\vec{m},\alpha}}
\label{skm}
\eea
for every layer $\alpha$. We call $\Psi^{corr}$ the $\Psi$ calculated from this ensemble.
\item {\it Uncorrelated weighted multiplex ensemble.}
In this ensemble, we set the expected degree $k_i^{\alpha}$ of every node $i$ in every layer $\alpha=1,2$ to be equal to the sum of the multidegrees (with $m_{\alpha}=1$) in the correlated weighted multiplex ensemble. We set the expected strengths $s_i^{\alpha}$ of every node $i$ in every layer $\alpha$ to be equal to the sum of the multistrengths of node $i$ in layer $\alpha$ in the correlated weighted multiplex ensemble. We call $\Psi^{corr}$ the $\Psi$ calculated from this ensemble.
\end{itemize}

In the correlated weighted multiplex ensemble the properties of the multilinks are accounted for, while in the uncorrelated weighted multiplex ensemble the different layers of the multiplex networks are analyzed separately (see Supplementary Material for the details). Finally, to quantify the additional amount of information carried by the correlated multiplex ensemble with respect to the uncorrelated multiplex ensemble, we define the indicator $\Xi$ as
\bea
\Xi=\frac{\Psi^{corr}}{\Psi^{uncorr}}.
\label{xi}
\eea 

As an example of a possible application of the indicator $\Xi$, we focus on a case inspired by the CoCi-PRE multiplex network, where we consider different exponents $\beta_{\vec{m},\alpha,in/out}$ for different multilinks. First, we created the correlated multiplex ensemble with power-law multidegree distributions with exponents $\gamma_{(1,m_2)}=2.6$ for $m_2=0,1$ and $\gamma_{(0,1),(in/out)}=1.9$, where  (for multidegree $(0,1)$ we imposed a structural cut-off). 
Multistrengths satisfy Eq.~(\ref{skm}), with $c_{\vec{m},\alpha}=1$ and $\beta_{(1,m_2),1}=1,$ for $m_2=0,1$; $\beta_{(1,1), 2}=1.3$, $\beta_{(0,1),2}=1.1$. Second, for the second layer, we created the uncorrelated version of the multiplex ensemble which is characterized by a super-linear dependence of the average strength on the  degree of the nodes. 
 We then measured $\Psi$ as a function of network size $N$ for these different ensembles. Numerically, the average $\Avg{\ldots}_{\pi(w)}$ was evaluated from $100$ randomizations. Figure~\ref{fig4} shows that $\Psi$ increases with network size $N$ as a power law, and that $\Xi$ fluctuates around an average value of $1.256$. These findings indicate that a significant amount of information is contained in multistrength and cannot be extracted from individual layers separately. Similar results, not shown here, were obtained with a correlated weighted multiplex ensemble characterized by non-trivial inverse multiparticipation ratios.
 
 \begin{figure}
  \begin{center}
   \centerline{\includegraphics[width=3.0in]{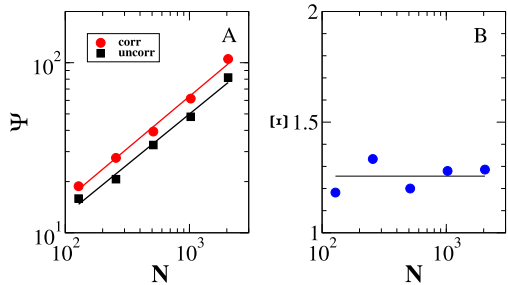}}
  \end{center}
  \caption{\label{fig4} (A) Value of the indicator $\Psi$ defined in Eq. $(\ref{psi})$ indicating the amount of information carried by the correlated and the uncorrelated multiplex ensembles  of $N$ nodes with respect to a null model in which the weights are distributed uniformly over the multiplex network. (B) Value of the indicator $\Xi$ defined in Eq. $(\ref{xi})$ indicating the additional amount of information encoded in the properties of multilinks in the correlated multiplex ensemble with respect to the corresponding uncorrelated multiplex ensemble. The solid line  refers to the average value  of $\Xi$ over the different multiplex network sizes.}
\end{figure}

\section{Conclusions}
In this paper, we have shown that weighted multiplex networks are characterized by significant correlations across layers, and in particular that weights are closely correlated with the multiplex network structure. To properly detect these correlations, we have introduced and defined two novel weighted properties of multiplex networks, namely the multistrength and the inverse multiparticipation ratio, that cannot be reduced to the properties of single layers. These weighted multiplex properties capture the crucial role played by multilinks in the distribution of weights, i.e. the extent to which there is a link connecting each pair of nodes in every layer of the multiplex network. To illustrate an example of weighted multiplex networks displaying non-trivial correlations between weights and topology, we analyzed the weighted properties of multilinks in two multiplex networks constructed by combining the co-authorship and citation networks involving the authors included in the APS dataset. Finally, based on the entropy of multiplex ensembles, we developed a theoretical framework for evaluating the information encoded in weighted multiplex networks, and proposed the indicator $\Psi$ for quantifying the information that can be extracted from a given dataset with respect to a null model in which weights are randomly distributed across links. Moreover, we have proposed a new indicator $\Xi$ that can be used to evaluate the additional amount of information that the weighted properties of multilinks provide over the information contained in the properties of single layers. In summary, in this paper we have provided compelling evidence that the analysis of multiplex networks cannot be simplified to the partial analysis of single layers, and in particular that non-trivial information can be uncovered only by shifting emphasis on a number of weighted properties of multilinks.



\section{Materials and Methods}
We can build a multiplex ensemble by maximizing the entropy ${\cal S}$ of the ensemble given by Eq. $(\ref{entropy})$ under the condition that the constraints imposed upon the multiplex networks are satisfied on average over the ensemble (soft constraints). We assume there are $K$ of such constraints determined by the conditions
\begin{equation}
\sum_{\vec{G}}P(\vec{G})F_{\mu}(\vec{G})=C_{\mu},
\label{constraints}
\end{equation}
for $\mu=1,2\ldots, K$, where $F_{\mu}(\vec{G})$ determines one of the structural constraints that we want to impose on average on the multiplex network. The most unbiased multiplex ensemble satisfying the constraints given by Eqs. $(\ref{constraints})$ maximizes the entropy ${\cal S}$ under these constraints. In this ensemble, the probability $P(\vec{G})$ for a multiplex network $\vec{G}$ of the ensemble is given by 
\begin{equation}
P(\vec{G})=\frac{1}{Z}\exp\left[-\sum_{\mu}\omega_{\mu}F_{\mu}(\vec{G})\right],
\label{PC}
\end{equation}
where the normalization constant $Z$ is called the ``partition function" of the canonical multiplex ensemble and is fixed by the normalization condition imposed on $P(\vec{G})$, whereas $\omega_{\mu}$ are the Lagrangian multipliers enforcing the constraints in Eq. $(\ref{constraints})$. The values of the Lagrangian multipliers $\omega_{\mu}$ are determined by imposing the constraints given by Eq. $(\ref{constraints})$, while for the probability $P(\vec{G})$ the structural form given by Eq. $(\ref{PC})$ is assumed. We refer to the entropy ${\cal S}$ given by Eq. $(\ref{entropy})$ calculated using the probability $P(\vec{G})$ given by Eq. $(\ref{PC})$ as the Shannon entropy of the multiplex ensemble. For all the details on the derivation of the entropy for these ensembles, we refer the interested reader to the Supplementary Material.


G. M. acknowledges the kind hospitality of Queen Mary University of London. G. M. and D. R. acknowledge EU FP7 MIMOmics Grant n. 305280.






\appendix
\clearpage

\onecolumngrid

\section{Additional information on the multiplex networks analyzed in this study}
\label{uno}
\subsection{The details on the two datasets}
We have considered the American Physical Society (APS) research data that is organized into two main datasets:
\begin{itemize}
\item {\em Article metadata:} for each article the metadata includes DOI, journal, volume, issue, first page and last page, article id and number of pages, title, authors, affiliations, publication history, PACS codes, table of contents, heading, article type, and copyright information.
\item {\em Citing article pairs:} this dataset consists of pairs of APS articles that cite each other. Each pair is represented by a pair of DOIs. The first id cites the second id.
\end{itemize}
In the APS metadata an author is usually identified by given name, middle name, and surname. In different articles, the same author can appear with her full name or with her initials. To deal with this issue, we decided to identify a specific author with the initials of his/her given name and middle name and with his/her full surname.\\
We restricted our analysis to the article metadata and citing article pairs that relate only to PRL and PRE. The total number of PRL articles is $95,516$ and the total number of PRL authors is $117,412$. The total number of PRE articles is $35,944$ and the total number of PRE authors is $36,171$. The number of authors that published both in PRE and PRL is equal to $17,470$.\\

Among the papers published in PRE and PRL, we focused our study only on those containing a number of authors $n_p\leq 10$. This excludes most of the experimental high-energy collaborations that are typically characterized by a number of authors of a different order of magnitude. We decided to place such a cut-off to the maximum number of authors allowed per paper to avoid biases due to very large publications. Given the cut-off, our study thus becomes limited to $35,766$ PRE articles (99.5 \%) and $35,205$ PRE authors (97.3 \%) on the one hand, and $89,245$ PRL articles (93.4 \%) and $92,436$ PRL authors (78.7 \%) on the other. The intersection of these two datasets includes $16,207$ authors (i.e., 92.8 \% of the previous intersection).\\

We analyzed two types of interaction between APS authors: scientific collaborations and citations, with weights defined as follows.
\begin{itemize}
\item {\em Collaborations:} two authors are connected if they co-authored at least one paper. The collaborative interaction between author $i$ and author $j$ is defined as in \cite{Newman2001,Barrat}, i.e. the undirected adjacency matrix element $a_{ij}$ is given by 
\begin{eqnarray}
a_{ij}&=&\sum_{p\in I}=\frac{\delta_{i}^p \delta_{j}^p}{n_p-1}\quad i \ne j\\
a_{ii}&=&0,
\end{eqnarray}
where the index $p$ indicates an article in the dataset $I$, $n_p$ indicates the number of authors of article $p$ and $\delta_{i}^p=1$ if node $i$ is an author of article $p$, and $\delta_{i}^p=0$ otherwise. The resulting network is undirected and without self-loops.

\item {\em Citations:} two authors are connected by a directed link if one author cites the other one. In this case, the element $a_{ij}$ of the directed adjacency matrix indicating how many times node $i$ cites node $j$ is given by 
\begin{eqnarray}
a_{ij}=\sum_{p,p'\in I }\delta_{i}^p\delta_{j}^{p'}b_{p,p'},
\end{eqnarray}
where $b_{p,p'}=1$ if article $p$ cites article $p'$, and $b_{p,p'}=0$ otherwise. Moreover $\delta_{i}^p$ is defined as above and indicates whether $i$ is author of article $p$ ($\delta_i^p=1$) or not ($\delta_{i}^p=0$). The resulting network is directed and with self-loops.
\end{itemize}

We constructed the following two duplex networks:
\begin{enumerate}
 \item
{\bf CoCo-PRL/PRE: }{\em collaborations among PRL and PRE authors}
The nodes of this multiplex network are the authors who published articles both in PRL and PRE (i.e., $16,207$ authors).
These nodes are connected in layer 1 through weighted undirected links indicating the strength of their collaboration in PRL (i.e., co-authorship of PRL articles).
The same nodes are connected in layer 2 through weighted undirected links indicating the strength of their collaborations in PRE (i.e., co-authorship of PRE articles).
\item
{\bf CoCi-PRE: }{\em collaborations among PRE authors and citations to PRE articles}
The nodes of this multiplex network are the authors of articles published in PRE ((i.e., $35,205$ authors). These nodes are connected in layer 1 through weighted undirected links indicating the strength of their collaboration in PRE (i.e., co-authorship of PRE articles). The same nodes are connected in layer 2 through weighted directed links indicating how many times an author (with articles in PRE) cited another author's work, where citations are limited to those made to PRE articles.
\end{enumerate}

\subsection{Total overlap and total weighted overlap of the multiplex networks}
In order to characterize the overlap existing between the links of the multiplex networks, we define the total overlap $O^{\alpha,\alpha'}$ between layer $\alpha$ and layer $\alpha'$
as the total number of pair of nodes $(i,j)$ connected both in layer $\alpha$ and in layer $\alpha'$, i.e.
\begin{eqnarray}
O^{\alpha,\alpha'}=\sum_{i<j} \theta(a_{ij}^{\alpha})\theta(a_{ij}^{\alpha'}),
\end{eqnarray}
where $\theta(x)=1$ if $x>1$ and $\theta(x)=0$ otherwise.
This definition can be extended to weighted multiplex networks by defining the {\em total weighted overlap} $O^{(w)\alpha,\alpha'}$ between layer $\alpha$ and layer $\alpha'$ as
\begin{eqnarray}
O^{(w),\alpha,\alpha'}=\sum_{i<j}\min\left( \frac{w_{ij}^{\alpha}}{w_{max}^{\alpha}}\frac{w_{ij}^{\alpha'}}{w_{max}^{\alpha'}}\right),
\end{eqnarray}
where $w_{max}^{\alpha}$ is the maximal weight in layer $\alpha$.
Table $\ref{table1}$ reports details on the total overlap and total weighted overlap, and indeed shows that our multiplex networks are characterized by a significant overlap of links.

 \begin{table}[t!]
\caption{Total overlap and total weighted overlap in the CoCo-PRL/PRE and CoCi-PRE multiplex networks.}
 \begin{tabular}{@{\vrule height 10pt depth20pt  width1pt}cccc|}
\hline
{\bf Dataset} & {\bf Layer} & {\bf Total overlap} \% &{\bf Total weighted overlap} \%\\
\hline
{ CoCo-PRL/PRE} & PRL      & 35.75 & 28.35\\
{ CoCo-PRL/PRE}& PRE      & 39.10  & 33.84\\
\hline
\hline
{ CoCi-PRE} &coll      & 39.51  & 14.24\\
{ CoCi-PRE}& cit      & 12.64  & 20.76\\
\hline
\end{tabular}
\label{table1}
\end{table}

 \subsection{Degree and multidegree distribution of the two multiplex networks}
The nodes $i=1,2\ldots, N$ of the multiplex networks have degrees $k^1_i$ in layer 1 and $k_i^2$ in layer 2. Moreover, we can define the multidegree $k_i^{\vec{m}}$ of a generic node $i$ as the sum of the multilinks $\vec{m}$ incident on it.
We observe that, since we always have 
\bea
k^{\vec{0}}_i=(N-1)-\sum_{\vec{m}\neq 0}k^{\vec{m}}_i,
\eea
we can therefore restrict the analysis to multidegrees $\vec{m}\neq \vec{0}$. Figures $\ref{Pk1}$ and $\ref{Pk2}$ show that the degree and multidegree for both the CoCo-PRL/PRE and the CoCi-PRE multiplex network are broadly distributed. In particular, we fitted the distributions with a power law and obtained exponents $\gamma$ indicated in Tables $\ref{gamma_exp1}$ and $\ref{gamma_exp2}$. Moreover, in both duplex networks, the degrees each author has in the two layers are positively correlated, as indicated by the Pearson correlation coefficient between degrees (See Tables $\ref{table_rho1}$, $\ref{table_rho2}$). Finally, also multidegrees in the multiplex networks are correlated, as indicated by their Pearson coefficients (See Tables $\ref{table_rho1multidegree}$, $\ref{table_rho2multidegree}$).

\begin{table}[h!]
\caption{  Fitted power-law exponents $\gamma\pm\Delta \gamma$ of the degree and multidegree distributions in the CoCo-PRL/PRE multiplex network.}
\begin{tabular}{@{\vrule height 10pt depth20pt  width1pt}c|cc||ccc|}
\hline
 & $P(k_1)$  & $P(k_2)$   & $P(k_{11})$   & $P(k_{10})$ & $P(k_{01})$\\
\hline
$\gamma$      & 2.50  & 2.62  & 2.90   &  2.51  & 2.64  \\
$\Delta \gamma$      &  0.14 & 0.16 &  0.30  &   0.13 &0.17 \\
\hline
\hline
\end{tabular}
\label{gamma_exp1}
\end{table}

\begin{table}[h!]
\caption{ Fitted power-law exponents $\gamma\pm \Delta \gamma$ of the degree and multidegree distributions in the CoCi-PRE multiplex network.}
\begin{tabular}{@{\vrule height 10pt depth20pt  width1pt}c|ccc||cccccc|}
\hline
 & $P(k_1)$  & $P(k_2^{in})$ & $P(k_2^{out})$  & $P(k_{11}^{in})$  & $P(k_{11}^{out})$ & $P(k_{10}^{in})$ & $P(k_{10}^{out})$& $P(k_{01}^{in})$& $P(k_{01}^{out})$\\
\hline
 $\gamma$     & 2.63  & 1.89  & 2.15&  2.58 & 2.57 & 3.02 & 2.99 & 1.89 & 2.18 \\
  $\Delta \gamma$     &  0.14 &0.06 &0.07&   0.20 &  0.17 & 0.23 & 0.21 & 0.06 &  0.07 \\
\hline
\hline
\end{tabular}
\label{gamma_exp2}
\end{table}

\begin{figure}
\begin{center}
\centerline{\includegraphics[width=7in]{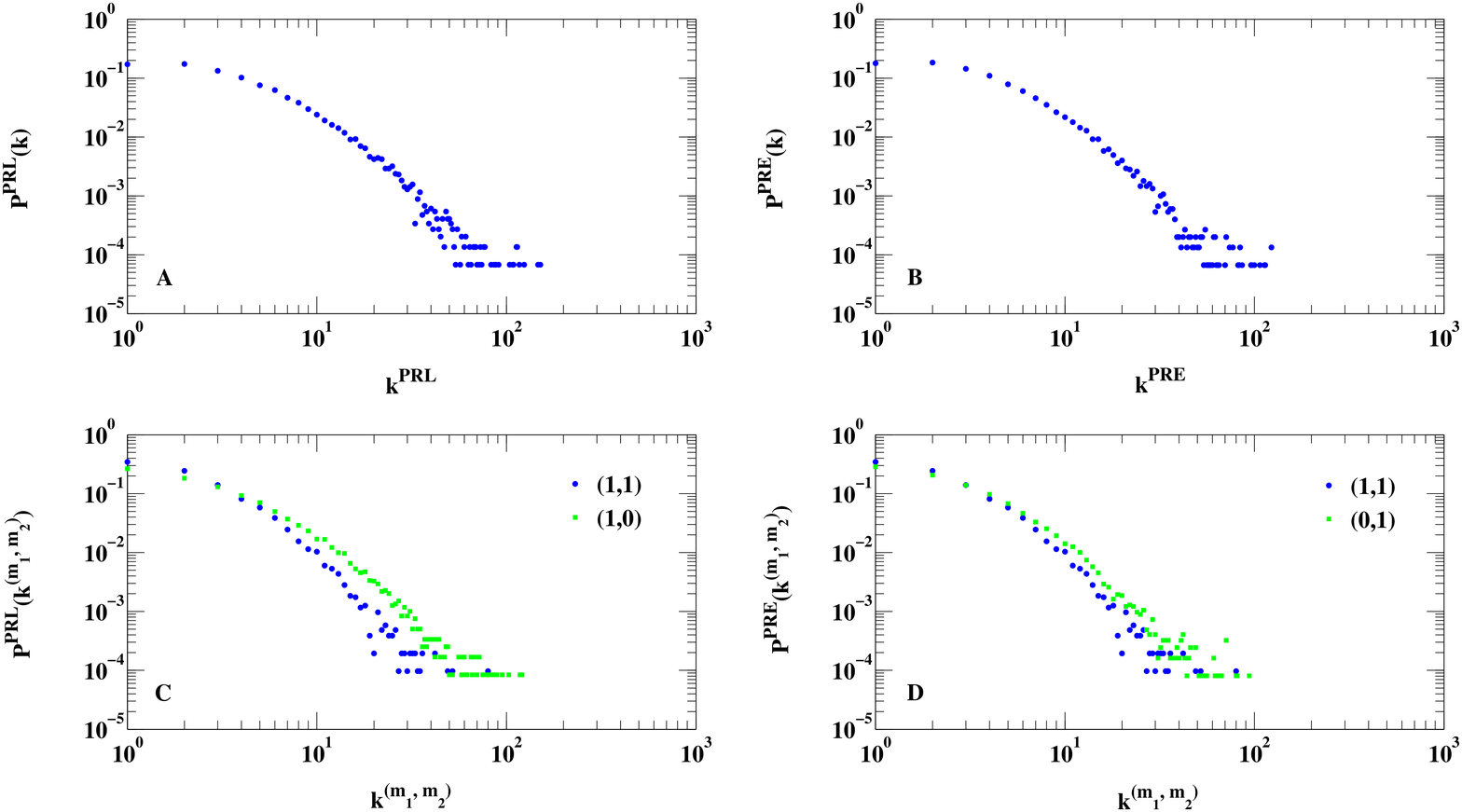}    }
\end{center}
\caption{The degree distributions and the multidegree distributions in the CoCo-PRL/PRE multiplex network. Related exponents in Tab. \ref{gamma_exp1}.}
\label{Pk1}
\end{figure}


\begin{figure}
\begin{center}
\centerline{\includegraphics[width=7in]{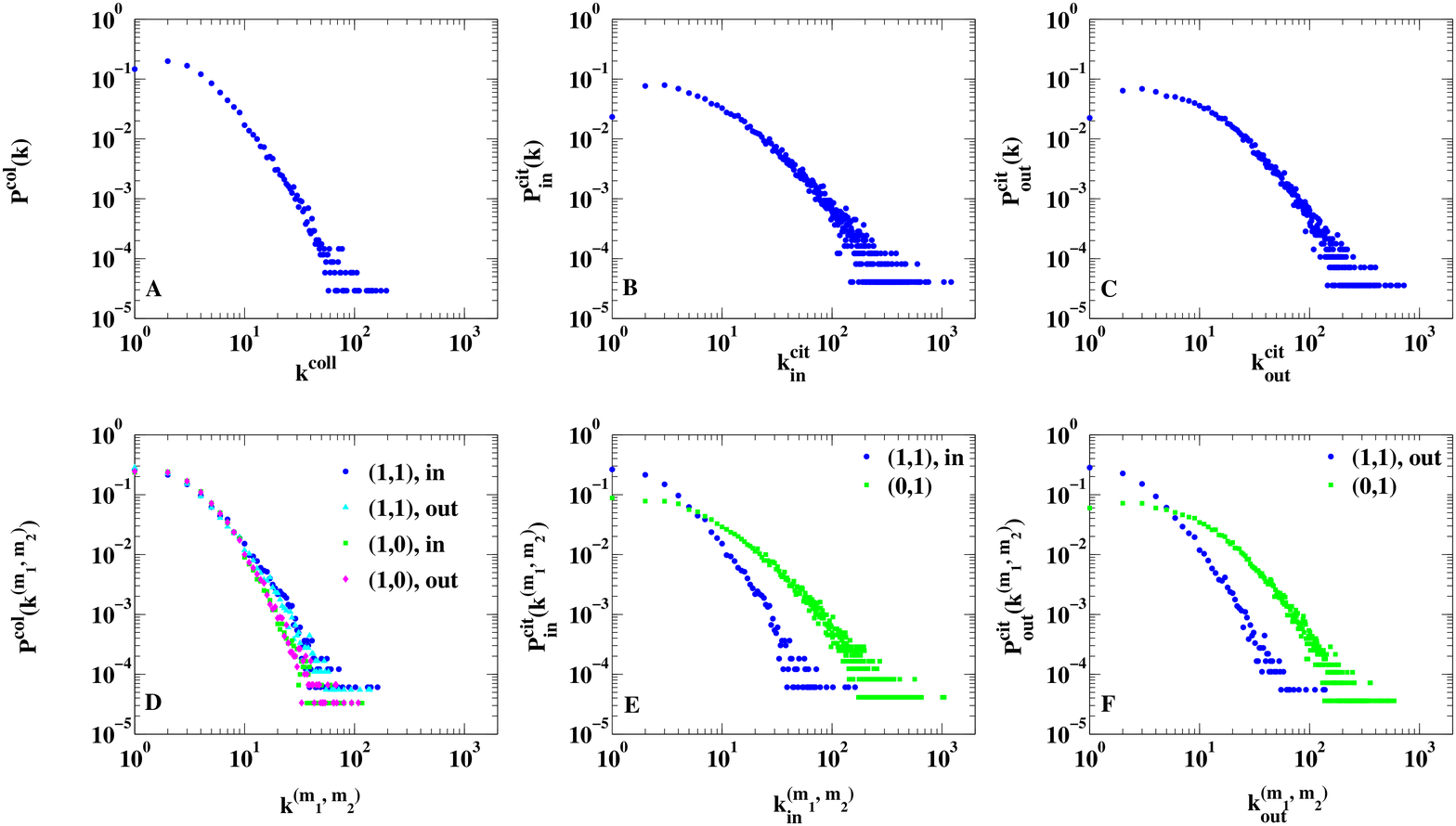}    }
\end{center}
\caption{The degree distributions and the multidegree distributions in the CoCi-PRE multiplex network. Related exponents in Tab. \ref{gamma_exp2}.}
\label{Pk2}
\end{figure}

\begin{table}[h!]
\caption{Pearson coefficients $\rho$ measuring the correlations between the degrees in the different layers and the strengths in the different layers in the CoCi-PRE multiplex network.}
\begin{tabular}{@{\vrule height 10pt depth20pt  width1pt}cccc||c ccc|}
\hline
$\rho$ & $k_1$  & $k_2^{in}$ & $k_2^{out}$ &$\rho$ & $s_1$  & $s_2^{in}$ & $s_2^{out}$\\
\hline
$k_1$      & 1 & 0.70 & 0.79 &$s_1$      & 1 & 0.80 & 0.88 \\
$k_2^{in}$ & 0.70 & 1 & 0.71 & $s_2^{in}$ & 0.80 & 1 & 0.80\\
$k_2^{out}$ & 0.79 & 0.71 & 1 &$s_2^{out}$ & 0.88 & 0.80 & 1 \\
\hline
\hline
\end{tabular}
\label{table_rho1}
\end{table}

\begin{table}[h!]
\caption{Pearson coefficients $\rho$ measuring the correlations between the degrees in the different layers and the strengths in the different layers in the CoCo-PRL/PRE multiplex network.}
\begin{tabular}{@{\vrule height 10pt depth20pt  width1pt}ccc||ccc|}
\hline
$\rho$ & $k_1$  & $k_2$ &$\rho$ & $s_1$  & $s_2$\\
\hline
$k_1$      & 1 & 0.74&$s_1$      & 1 & 0.65\\
$k_2$ & 0.74 & 1 &$s_2$ & 0.65 & 1\\
\hline
\hline
\end{tabular}
\label{table_rho2}
\end{table}


\begin{table}[h!]
\caption{Pearson coefficients $\rho$ measuring the correlations between multidegrees in the CoCi-PRE multiplex network.}
\begin{tabular}{@{\vrule height 10pt depth20pt  width1pt}ccccccc|}
\hline
$\rho$ & $k_{11}^{in}$  & $k_{11}^{out}$ & $k_{10}^{in}$ & $k_{10}^{out}$ & $k_{01}^{in}$  & $k_{01}^{out}$\\
\hline
$k_{11}^{in}$ &   1 &   0.93  & 0.36  &0.52  &  0.71  &  0.72 \\
$k_{11}^{out}$ &    0.93  &  1  & 0.46 & 0.45 &  0.67  &  0.75 \\
$k_{10}^{in}$  &   0.36  &  0.46  &  1 & 0.90  &  0.32  &  0.48 \\
$k_{10}^{out}$&   0.52 &    0.45  & 0.90 & 1 &    0.43 &    0.51 \\
$k_{01}^{in}$ &    0.71 &    0.67 & 0.32 &  0.43 &    1 &    0.65\\
$k_{01}^{out}$ &   0.72 &    0.75 &  0.48 & 0.51 &   0.65  &    1\\
\hline
\hline
\end{tabular}
\label{table_rho1multidegree}
\end{table}

\begin{table}[h!]
\caption{Pearson coefficients $\rho$ measuring the correlations between multidegrees in the CoCo-PRL/PRE multiplex network.}
\begin{tabular}{@{\vrule height 10pt depth20pt  width1pt}cccc|}
\hline
$\rho$ & $k_{11}$ & $k_{10}$  & $k_{01}$\\
\hline
$k_{11}$  &  1  &  0.41 &   0.46\\
$k_{10}$  &  0.41 &   1  &  0.53\\
$k_{01}$  &  0.46  &  0.53  & 1\\
\hline
\hline
\end{tabular}
\label{table_rho2multidegree}
\end{table}

\subsection{Weighted network  properties of single layers}
Here we report the weighted network properties of the single layers of our multiplex networks. In general, the average strength $s_k^{\alpha} $of nodes with degree $k$ in layer $\alpha$ and the average inverse participation ratio $Y_k^{\alpha}$ of nodes with degree $k$ in layer $\alpha$ are described by the functional behavior
\bea
s_k^{\alpha}\propto k^{\beta_{\alpha}},\nonumber \\
Y_k^{\alpha}\propto \frac{1}{k^{\lambda_{\alpha}}}.
\label{Sksingle}
\eea
As shown by Figure $\ref{figsk1}$, the CoCo-PRL/PRE multiplex network is characterized by a linear behavior of average strength as a function of the degree of nodes, i.e.
$\beta_1=0.96\pm 0.04$, $\beta_2=1.01\pm 0.05$, where the first layer indicates the {\it PRL} collaboration network and the second layer the {\it PRE} collaboration network. The exponents $\lambda_{\alpha}$ in the CoCo-PRL/PRE multiplex network are given by $\lambda_1=0.84\pm 0.03$ and $\lambda_2=0.80\pm 0.05$. Figure $\ref{figsk2}$ shows that the CoCi-PRE multiplex network is characterized by a linear behavior of average strength as a function of the degree of nodes in the collaboration network ($\beta_1=1.03\pm 0.04$), and by a super-linear behavior in the citation network, i.e. $\beta_2^{in}=1.13\pm 0.02$, $\beta_2^{out}=1.14\pm 0.03$, where the first layer indicates the {\it PRE} collaboration network and the second layer the directed citation network. The exponents $\lambda_{\alpha}$ in the CoCi-PRE multiplex network are given by $\lambda_1=0.79\pm 0.04$ and $\lambda_2^{in}=0.72\pm 0.03$, $\lambda_2^{out}=0.70\pm 0.04$.


\begin{figure}[h!]
\begin{center}
\centerline{\includegraphics[width=8in]{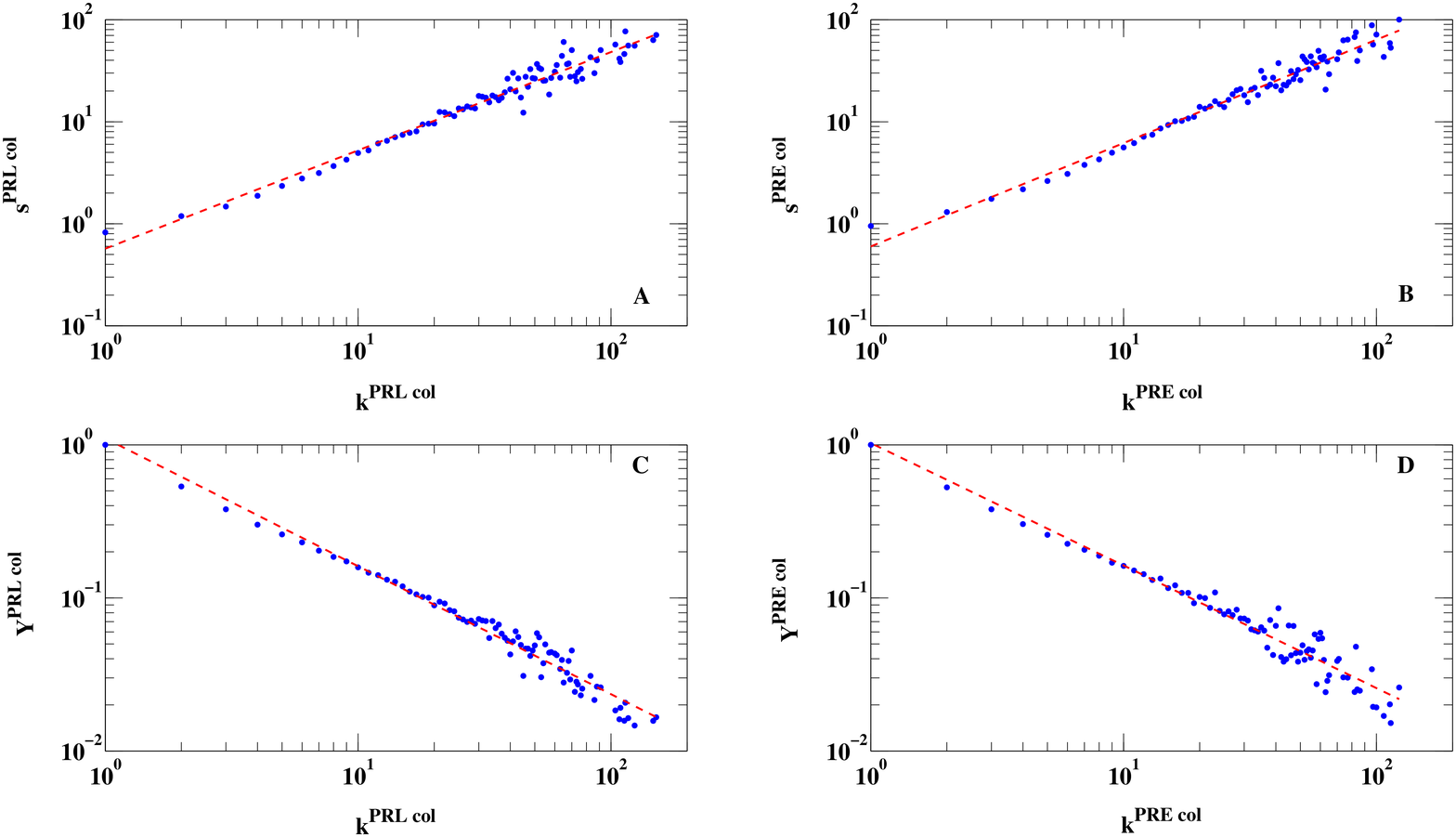}}
 \end{center}
  \caption{Average strength versus degree and average inverse participation ratio in the two layers of the CoCo-PRL/PRE multiplex network. The average strengths and the average inverse participation ratio follow the functional form described by Eq. $(\ref{Sksingle})$.}
  \label{figsk1}
\end{figure}
\begin{figure*}[h!]
  \begin{center}
 \includegraphics[width=7.5in]{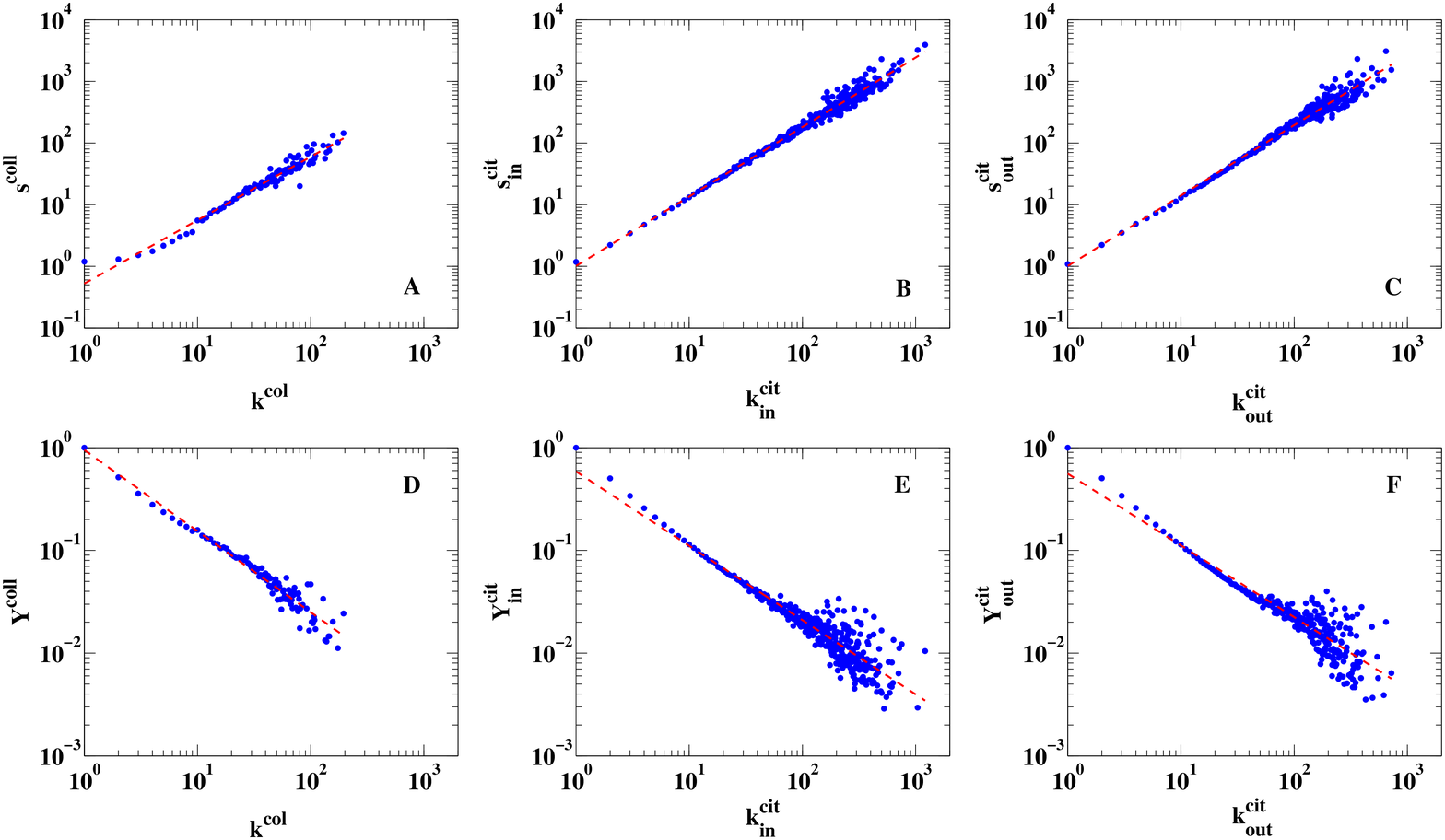}    
  \end{center}
  \caption{Average strength and average inverse participation ratio versus degree in the two layers of the CoCi-PRE multiplex network. Average strength and average inverse participation ratio follow the functional form described by Eq. $(\ref{Sksingle})$. }
  \label{figsk2}
\end{figure*}

\subsection{Statistical analysis of  the properties of multilinks in the CoCo-PRL/PRE multiplex network}
In this subsection we discuss in detail the results of our statistical analysis of the properties of multilinks in the CoCo-PRL/PRE multiplex network. In particular, we focus on the average multistrength of nodes with a given multidegree, i.e $s^{\vec{m},\alpha}(k^{\vec{m}})=\Avg{s_i^{\vec{m},\alpha}\delta(k_i^{\vec{m}},k^{\vec{m}})}$, and the average inverse multiparticipation ratio of nodes with a given multidegree, $Y^{\vec{m},\alpha}(k^{\vec{m}})=\Avg{Y_i^{\vec{m},\alpha}\delta(k_i^{\vec{m}},k^{\vec{m}})}$. These quantities are expected to scale as 
\bea
s^{\vec{m},\alpha}(k^{\vec{m}}) &=&e^{q^{\vec{m},\alpha}}(k^{\vec{m}})^{\beta_{\vec{m},\alpha}} \nonumber \\
Y^{\vec{m},\alpha}(k^{\vec{m}})&= &e^{p^{\vec{m},\alpha}}\frac{1}{(k^{\vec{m}})^{\lambda_{\vec{m},\alpha}}},
\label{SkYk}
\eea
with exponents $\beta_{\vec{m},\alpha} \geq 1$ and $\lambda_{\vec{m},\alpha}\leq 1$.
In what follows, we will label the {\it PRL} collaboration layer as $\alpha=1$ and the {\it PRE} collaboration layer as $\alpha=2$ .\\

\subsubsection{Statistical analysis of the average multistrengths in the CoCo-PRL/PRE multiplex network}
In the CoCo-PRL/PRE multiplex network {(see Fig. 2 in the main text)}, the fitted exponents $\beta_{\vec{m},1}$ for multilinks $\vec{m}=(1,1)$ and $\vec{m}=(1,0)$ are not significantly different ($\beta_{(1,1),1}=1.06 \pm 0.09$, $\beta_{(1,0),1}=0.97 \pm 0.03$). Moreover, also the fitted proportionality constants in Eq. $(\ref{SkYk})$ are not significantly different, with values $q_{(1,1),1}=-0.51 \pm 0.26$ and $q_{(1,0),1}=-0.78 \pm 0.10$. However, we can perform a paired samples Student's t-test to show how the average multistrength per fixed multidegree $s^{\vec{m},\alpha}(k^{\vec{m}})$ is significantly higher for multilinks $(1,1)$ than multilinks $(1,0)$. We have identified pairs of average multistrength $s^{(1,1),1}(k^{(1,1)})$ and $s_k^{(1,0),1}(k^{(1,0)})$, corresponding to the same multidegree value $k^{(1,1)}=k^{(1,0)}=k$. The paired samples Student's t-test returns a test decision for the null hypothesis that the values $\log\left(s^{(1,1),1}(k)/s^{(1,0),1}(k)\right)$ come from a normal distribution with mean zero and variance from the data. In our case, the null hypothesis is rejected with a p-value equal to $2.90\cdot 10^{-16}$. Furthermore, $\Avg{\log\left(s^{(1,1),1}(k)/s^{(1,0),1}(k)\right)}$ is equal to 0.53. This analysis suggests that for a particular value of degree $k$ the related $s^{(1,1),1}(k)$ is higher than $s^{(1,0),1}(k)$ and the two average multistrengths satisfy the relation $s^{(1,1),1}(k)\approx e^{0.53} s^{(1,0),1}(k)$. Similar results were obtained in the case of the multistrengths on the second layer indicating the collaboration network on {\it PRE} articles. The null hypothesis is rejected with a p-value equal to $8.98\cdot 10^{-15}$ and $\Avg{\log\left(s^{(1,1),2}(k)/s^{(0,1),2}(k)\right)}$ is equal to $0.57$.

\subsubsection{Statistical analysis of the average inverse multiparticipation ratio in the CoCo-PRL/PRE multiplex network}
In the {\it PRL} layer the fitted exponents $\lambda_{\vec{m}, \alpha}$ are significantly different and their values are $\lambda_{(1,1),1}=0.74\pm0.05$ and $\lambda_{(1,0),1}=0.84\pm 0.03$. The weights regarding multilinks $(1,1)$ are therefore distributed more heterogeneously than the weights regarding multilinks $(1,0)$. Similarly, in the {\it PRE} layer, the fitted exponents are $\lambda_{(1,1),2}=0.73\pm0.06$ and $\lambda_{(0,1),2}=0.84\pm 0.05$. 

The paired Student's t-test is also useful to understand the properties of the average inverse multiparticipation ratio. In addition to the fitted exponents, we can perform a t-test as we did previously considering now $Y^{\vec{m},\alpha}(k^{\vec{m}})$. This test underlines how the inverse multiparticipation ratios regarding multilinks (1,1) are significantly higher than those regarding multilinks (1,0) or (0,1).
In the case $Y^{(1,1),1}(k)$ vs $Y^{(1,0),1}(k)$, the t-test gives a p-value equal to $0.002$ and an average value $\Avg{\log\left(Y^{(1,1),1}(k)/Y^{(1,0),1}(k)\right)}=0.11$. In the case $Y^{(1,1),2}$ vs $Y^{(0,1),2}(k)$, the p-value is equal to $6.64\cdot 10^{-6}$, and the average value is $\Avg{\log\left(Y^{(1,1),2}(k)/Y^{(0,1),2}(k)\right)}=0.19$.

\subsection{Statistical analysis of the properties of multilinks in the CoCi-PRE multiplex network}
We analyzed the average multistrength of nodes with a given multidegree, i.e $s^{\vec{m},\alpha,(in/out)}(k^{\vec{m}(in/out)})=\Avg{s_i^{\vec{m},\alpha,(in/out)}\delta(k_i^{\vec{m},(in/out)},k^{\vec{m},(in/out)})}$, and the average inverse multiparticipation ratio of nodes with a given multidegree, $Y^{\vec{m},\alpha, (in/out)}(k^{\vec{m},(in/out)})=\Avg{Y_i^{\vec{m},\alpha,(in/out)}\delta(k_i^{\vec{m},(in/out)},k^{\vec{m},(in/out)})}$, where a distinction was made between incoming and outgoing links in the citation layer. These quantities are expected to scale as 
\bea
s^{\vec{m},1, (in, out)}(k^{\vec{m, (in, out)}}) &=&e^{q^{\vec{m, (in, out)},1}}(k^{\vec{m, (in, out)}})^{\beta_{\vec{m},1, (in, out)}} \nonumber \\
s^{\vec{m},2,(in/out)}(k^{\vec{m},(in/out)}) &=&e^{q^{\vec{m},2,(in/out)}}(k^{\vec{m},(in/out)})^{\beta_{\vec{m},2,(in/out)}} \nonumber \\
Y^{\vec{m},1,(in/out)}(k^{\vec{m},(in/out)})&= &e^{p^{\vec{m},1,(in/out)}}\frac{1}{(k^{\vec{m},(in/out)})^{\lambda_{\vec{m},1,(in/out)}}} \nonumber \\
Y^{\vec{m},2,(in/out)}(k^{\vec{m},(in/out)})&= &e^{p^{\vec{m},2,(in/out)}}\frac{1}{(k^{\vec{m},(in/out)})^{\lambda_{\vec{m},2,(in/out)}}},
\label{SkYk2}
\eea
with exponents $\beta_{\vec{m},\alpha,(in/out)} \geq 1$ and $\lambda_{\vec{m},\alpha, (in/out)}\leq 1$.
In what follows we will label the {\it PRE} collaboration layer as $\alpha=1$ and the {\it PRE} citation layer as $\alpha=2$.\\

\subsubsection{The statistical analysis of the average multistrengths in the CoCi-PRE multiplex network}
In the CoCi-PRE multiplex network {(see Fig. 3 in the main text )} we can perform, at first, a statistical analysis of the multistrengths in the collaboration layer. The fitted exponents $\beta_{(1,1),1,in}$, $\beta_{(1,1),1,out}$, $\beta_{(1,0),1, in}$ and $\beta_{(1,0),1, out}$, are not significantly different ($\beta_{(1,1),1,in}=1.03 \pm 0.04$, $\beta_{(1,1),1,out}=1.05 \pm 0.04$ , $\beta_{(1,0),1, in}=0.98 \pm 0.05$, $\beta_{(1,0),1, out}=0.97 \pm 0.05$). Conversely, the fitted intercepts of the log-log plot, regarding multilinks $(1,1),{(in/out)}$ are significantly different from the intercept for multilinks $(1,0),{(in/out)}$, namely, $q^{(1,1),1,in}=-0.41 \pm 0.15$ and $q^{(1,0),1,in}=-0.97 \pm 0.17$, $q^{(1,1),1,out}=-0.38 \pm 0.14$ and $q^{(1,0),1,out}=-0.95 \pm 0.16$. From a paired samples Student's t-test, in the same way as we did for the average multistrengths in the CoCo-PRL/PRE multiplex network, we obtained that both $s^{(1,1),1,in}(k)$ and $s^{(1,1),1,out}(k)$ are significantly higher than $s^{(1,0),1, in}(k)$ and $s^{(1,0),1, out}(k)$. In the case $s^{(1,1),1,in}(k)$ vs $s^{(1,0),1,in}(k)$, we have a p-value equal to $4.86\cdot 10^{-33}$ and an average value $\Avg{\log\left(s^{(1,1),1,in}(k)/s^{(1,0),1,in}(k)\right)}=0.72$. In the case $s^{(1,1),1,out}(k)$ vs $s^{(1,0),1, out}(k)$, we have a p-value equal to $9.93\cdot 10^{-30}$ and an average value $\Avg{\log\left(s^{(1,1),1,out}(k)/s^{(1,0),1, out}(k)\right)}=0.80$. Based on the fitted parameters and the Student's t-test, the data suggest that both the multidegree for multilinks $(1,1)$ and multilinks $(1,0)$ have a linear relation with their own multistrengths in the collaboration layer, and that multistrengths (1,1) are related to multistrengths (1,0) by a multiplicative constant.

In the citation layer, the fitted exponents indicate a super-linear scaling and are significantly different. For the in-citations, we have $\beta_{(1,1),2,in}=1.30\pm 0.07$ and $\beta_{(0,1),2,in}=1.11\pm 0.01$. The intercepts are $q^{(1,1), 2,in}=0.47\pm 0.25$ and $q^{(0,1), 2,in}=-0.01\pm 0.07$. For the out-citations, we have $\beta_{(1,1),2,out}=1.32\pm 0.08$ and $\beta_{(0,1),2,out}=1.10\pm 0.02$. The intercepts are $q^{(1,1), 2,out}=0.45\pm 0.26$ and $q^{(0,1), 2,out}=0.06\pm 0.09$. 

\subsubsection{The statistical analysis of the multi inverse participation ratio in the CoCi-PRE multiplex network}
In the collaboration layer the fitted exponents are $\lambda_{(1,1),1,in}=0.80 \pm 0.06$, $\lambda_{(1,1),1,out}=0.77 \pm 0.05$, $\lambda_{(1,0),1,in}=0.88\pm 0.03$ and $\lambda_{(1,0),1,out}=0.90\pm 0.02$ (the confidence intervals of the fitted exponents $\lambda_{(1,1),1,out}$ and $\lambda_{(1,0),1,out}$ do not overlap for a narrow window). Performing the t-test as usual, we found that the inverse multiparticipation ratio for multilinks $(1,1)$ is always larger than the inverse multiparticipation ratio for multilinks $(1,0)$. In the case $Y^{(1,1),1,in}(k)$ vs $Y^{(1,0),1,in}(k)$, the t-test gives a p-value equal to $5.73\cdot 10^{-17}$ and an average value $\Avg{\log\left(Y^{(1,1),1,in}(k)/Y^{(1,0),1,in}(k)\right)}=0.48$. In the case $Y^{(1,1),1,out}$ vs $Y^{(1,0),1,out}(k)$, the p-value is equal to $5.48\cdot 10^{-19}$ and the average value is $\Avg{\log\left(Y^{(1,1),1,out}(k)/Y^{(1,0),1,out}(k)\right)}=0.33$.

In the $in-$ and $out-$citation layers, the fitted exponents $\lambda_{\vec{m},2,(in/out)}$ regarding multilinks $(1,1)$ are not significantly different from those regarding multilinks $(0,1)$ ($\lambda_{(1,1),2,in}=0.73\pm 0.05$, $\lambda_{(0,1),2,in}=0.74\pm 0.04$ and  $\lambda_{(1,1),2,out}=0.75\pm 0.05$, $\lambda_{(0,1),2,out}=0.69\pm 0.05$ ).
Nevertheless, the paired Student's t-test shows how the inverse multiparticipation ratio for multilinks $(1,1)$ is always larger than the inverse multiparticipation ratio for multilinks $(0,1)$. 
In the case $Y^{(1,1),2,in}(k)$ vs $Y^{(0,1),2,in}(k)$, the t-test gives a p-value equal to $7.60\cdot 10^{-21}$ and an average value $\Avg{\log\left(Y^{(1,1),2,in}(k)/Y^{(0,1),2,in}(k)\right)}=0.34$. In the case $Y^{(1,1),2,out}(k)$ vs $Y^{(0,1),2,out}(k)$, the p-value is equal to $1.12\cdot 10^{-15}$ and the average value is $\Avg{\log\left(Y^{(1,1),2,out}(k)/Y^{(0,1),2,out}(k)\right)}=0.34$.

\section{Weighted Multiplex Ensembles}
\label{due}
\subsection{Definition }
A weighted multiplex network is formed by $N$ nodes connected within $M$ weighted networks $G_{\alpha}=(V,E_{\alpha})$, with $\alpha=1,\ldots,M$ and $|V|=N$.Therefore we can represent a multiplex network as $\vec{G}=(G_1,G_2,\ldots, G_{\alpha},\ldots G_M)$. Each network $G_{\alpha}$ is fully described by the adjacency matrix of elements $a_{ij}^{\alpha}$, with $a_{ij}^{\alpha}=w_{ij}^{\alpha}>0$ if there is a link of weight $w_{ij}^{\alpha}$ between nodes $i$ and $j$ in layer $\alpha$, and $a_{ij}^{\alpha}=0$ otherwise. In what follows, in order to simplify the treatment of the weighted multiplex networks, we will assume that the weight of the link between any pair of nodes $(i,j)$, $a_{ij}^{\alpha}=w_{ij}^{\alpha}$, can only take integer values. This is not a major limitation because in a large number of weighted multiplex networks the weights of the links can be considered as multiples of a minimal weight.

\subsection{Canonical weighted multiplex ensembles or exponential weighted multiplex ensembles}
The canonical network ensembles (also known as exponential random graphs) are a very powerful tool for building null models of networks \cite{Newman1,Munoz,PNAS}. Here we generalize the formalism developed for unweighted multiplex  ensembles \cite{BianconiPRE} to take weighted multiplex ensembles into account.

The construction of the canonical weighted multiplex ensembles or exponential random multiplex follows closely the derivation or the exponential random graphs.
A weighted multiplex ensemble is defined once the probability $P(\vec{G})$ of any possible weighted multiplex is given.
We can build a canonical multiplex ensemble by maximizing the entropy ${\cal S}$ of the ensemble given by 
\bea
{\cal S}=-\sum_{\vec{G}}P(\vec{G})\log P(\vec{G}),
\label{entropy}
\eea under the condition that the soft constraints we want to impose are satisfied.
We assume there are $K$ of such constraints determined by the conditions
\begin{equation}
\sum_{\vec{G}}P(\vec{G})F_{\mu}(\vec{G})=C_{\mu},
\label{constraints}
\end{equation}
for $\mu=1,2\ldots, K$, where $F_{\mu}(\vec{G})$ determines one of the  structural constraints that we want to impose on the multiplex network. 
Therefore the maximal-entropy multiplex ensemble satisfying the constraints given by Eqs. $(\ref{constraints})$ is the solution of the following system of equations 
\begin{equation}
\frac{\partial }{\partial P(\vec{G})}\left[{\cal S}-\sum_{\mu=1}^K \omega_{\mu} \sum_{\vec{G}}F_{\mu}(\vec{G})P(\vec{G})-\Lambda\sum_{\vec{G}}P(\vec{G})\right]=0,
\end{equation}
where the Lagrangian multiplier $\Lambda$ enforces the normalization of the $P(\vec{G})$ probability distribution, and the Lagrangian multiplier $\omega_{\mu}$ enforces the constraint $\mu$.
Therefore we obtain that the probability of a multiplex network $P(\vec{G})$ in a canonical multiplex ensemble is given by 
\begin{equation}
P(\vec{G})=\frac{1}{Z}\exp\left[-\sum_{\mu}\omega_{\mu}F_{\mu}(\vec{G})\right],
\label{PC}
\end{equation}
where the normalization constant $Z=\exp(1 + \Lambda)$ is called the ``partition function" of the canonical multiplex ensemble and is fixed by the normalization condition on $P(\vec{G})$. Thus, $Z$ is given by 
\bea
Z=\sum_{\vec{G}}\exp\left[-\sum_{\mu}\omega_{\mu}F_{\mu}(\vec{G})\right].
\eea
The values of the Lagrangian multipliers $\omega_{\mu}$ are determined by imposing the constraints given by Eq. $(\ref{constraints})$ assuming for the probability $P(\vec{G})$ the structural form given by Eq. $(\ref{PC})$.
From the definition of the partition function $Z$ and Eq. $(\ref{PC})$, it can be easily shown that the Lagrangian multipliers $\omega_{\mu}$ can be expressed as the solutions of the following set of equations
\bea
C_{\mu}=-\frac{\partial \log Z}{\partial \omega_{\mu}}.
\eea
In this ensemble, we can then relate the entropy ${\cal S}$ (given by Eq. $(\ref{entropy})$) to the canonical partition function $Z$, and we obtain
\begin{eqnarray}
{\cal S}&=&\sum_{\mu} \omega_{\mu} C_{\mu}+\log{Z}.
\label{Sc}
\end{eqnarray}
We call the entropy ${\cal S}$ of the canonical multiplex ensemble the {\it Shannon entropy} of the ensemble.

\subsection{Uncorrelated and correlated canonical multiplex ensembles}
Multiplex ensembles can be distinguished between uncorrelated and correlated ones \cite{BianconiPRE}. For uncorrelated multiplex ensembles, the probability of a multiplex network $P(\vec{G})$ is factorizable into the probability $P_{\alpha}(G_{\alpha})$ of each single network $G_{\alpha}$ in layer $\alpha$, i.e.
\begin{equation}
\label{Puncorr}
P(\vec{G}) =\prod_{\alpha=1}^{M}P_{\alpha}(G_{\alpha}).
\end{equation}
Therefore, the entropy ${\cal S}$ of any uncorrelated multiplex ensemble given by Eq. $(\ref{entropy})$ with $P(\vec{G})$ given by Eq. $(\ref{Puncorr})$ is additive in the number of layers, i.e.
\begin{equation}
\label{entropy_noncorr}
{\cal S}=\sum_{\alpha=1}^M {\cal S}_{\alpha}=-\sum_{\alpha=1}^MP_{\alpha}(G_{\alpha})\log P_{\alpha}(G_{\alpha}).
\end{equation}

For a canonical uncorrelated multiplex ensemble, $P(\vec{G})$ has to satisfy both Eq. (\ref{Puncorr}) and Eq. (\ref{PC}). Therefore, in order to have an uncorrelated multiplex ensemble, the functions $F_{\mu}(\vec{G})$ should be equal to a linear combination of constraints $f_{\mu, \alpha}(G_{\alpha})$ on the networks $G_{\alpha}$ in a single layer $\alpha$, i.e.,
\begin{equation}
F_{\mu}(\vec{G})=\sum_{\alpha}^{M}f_{\mu, \alpha}(G_{\alpha}).
\end{equation}
A special case of this type of constraints is given when each constraint depends on a single network $G_{\alpha}$ in layer $\alpha$. An example of this type of constraints will be discussed in the following subsection where we will focus on the important case in which the constraints are the strength sequence $\{s_i^{\alpha}\}$ in any layer $\alpha$, and the degree sequence $\{k_i^{\alpha}\}$ in any layer $\alpha$.

Moroever, we can define the marginal probability for a specific value of the element $a_{ij}^{\alpha}$
\begin{equation}
\label{marginals}
\pi_{ij}^{\alpha}(a_{ij}^{\alpha}=w)=\sum_{\vec{G}}P(\vec{G})\delta(a_{ij}^{\alpha}, w),
\end{equation}
where $\delta(x,y)$ stands for the Kronecker delta. The marginal probabilities $\pi_{ij}^{\alpha}(a_{ij}^{\alpha})$ sum up to one
\begin{equation}
\sum_{a_{ij}^{\alpha}=0}^{\infty}\pi_{ij}(a_{ij}^{\alpha})=1.
\end{equation}
We can also calculate the average weight $\avg{a_{ij}^{\alpha}}$ of links between nodes $i$ and $j$ as
\begin{equation} 
\Avg{a_{ij}^{\alpha}}=\sum_{\vec{G}}P(\vec{G})a_{ij}^{\alpha}=\sum_{a_{ij}^{\alpha}=0}^{\infty}a_{ij}^{\alpha}\pi_{ij}(a_{ij}^{\alpha})
\label{av_weight}
\end{equation}
In layer $\alpha$ a link between two nodes $i$ and $j$ exists with probability $p_{ij}^{\alpha}$, related with all the possible weights different from zero
\begin{equation}
\label{probability}
p_{ij}^{\alpha}=\sum_{\vec{G}}P(\vec{G}) \theta(a_{ij}^{\alpha})=\sum_{a_{ij}^{\alpha}\ne 0}^{\infty}\pi_{ij}^{\alpha}(a_{ij}^{\alpha}).
\end{equation}

\subsection{Multiplex ensemble with given expected strength sequence and degree sequence in each layer}
Here we consider the relevant example of the uncorrelated multiplex ensemble in which we fix the expected strength $s_i^{\alpha}$ and the expected degree $k_i^{\alpha}$ of every node $i$ in each layer $\alpha$. We have $K=M \cdot 2N$ constraints in the system. These constraints are given by
\bea
\sum_{\vec{G}}F_{i,\alpha}(\vec{G}) P(\vec{G})&=\sum_{\vec{G}}\left ( \sum_{j \ne i} a_{ij}^{\alpha}\right)P(\vec{G})=s_i^{\alpha}\nonumber \\
\sum_{\vec{G}}F_{i,\alpha}(\vec{G}) P(\vec{G})&=\sum_{\vec{G}}\left ( \sum_{j \ne i} \theta(a_{ij}^{\alpha})\right)P(\vec{G})=k_i^{\alpha},
\label{constraints1}
\eea
with $\alpha=1,2,\ldots, M$.
We introduce the Lagrangian multipliers $w_{i,\alpha}$ for the first set of $N\cdot M$ constraints and the Lagrangian multipliers $\omega_{i,\alpha}$ for the second set of $N\cdot M$ constraints. Therefore, the probability $P(\vec{G})$ of a multiplex network in this ensemble, in general given by Eq. $(\ref{PC})$, in this specific example is given by  
\begin{equation}\hspace*{-3mm}
P(\vec{G})=\frac{1}{Z}\exp \left [ -\sum_{\alpha=1}^M \sum_{i=1}^Nw_{i,\alpha} \sum_{j \ne i} a_{ij}^{\alpha}-\sum_{\alpha=1}^M \sum_{i=1}^N \omega_{i,\alpha} \sum_{j \ne i} \theta(a_{ij}^{\alpha})\right]\nonumber,
\end{equation}
where the partition function $Z$ can be expressed explicitly as
\bea
Z&=&\sum_{\vec{G}}\exp \left [ -\sum_{\alpha=1}^M \sum_{i=1}^N \sum_{j \ne i}\left(w_{i,\alpha}  a_{ij}^{\alpha}+\omega_{i,\alpha} \theta(a_{ij}^{\alpha})\right)\right]\nonumber\\
&=&\prod_{\alpha=1}^{M}\prod_{i<j}\left ( 1+ \frac{e^{-(\omega_{i,\alpha}+\omega_{j,\alpha})-(w_{i,\alpha}+w_{j,\alpha})}}{1-e^{-(w_{i,\alpha}+w_{j,\alpha})}}\right),
\label{Zuncorr}
\eea
and the Lagrangian multipliers are fixed by the conditions Eqs. $(\ref{constraints1})$.
From Eq. (\ref{marginals}) we write the marginal probabilities $\pi_{ij}^{\alpha}(a_{ij}^{\alpha})$ for this specific ensemble that are given by 
\bea
\pi_{ij}^{\alpha}(a_{ij}^{\alpha})
&=&\frac{e^{-(w_{i,\alpha}+w_{j,\alpha})a_{ij}^{\alpha}-(\omega_{i,\alpha}+\omega_{j,\alpha})\theta(a_{ij}^{\alpha})}(1-e^{-(w_{i,\alpha}+w_{j,\alpha})})}{ 1+ e^{-(w_{i,\alpha}+w_{j,\alpha})}(e^{-(\omega_{i,\alpha}+\omega_{j,\alpha})}-1)}.
\label{pib}
\eea
The average weight of the link $(i,j)$ in layer $\alpha$, i.e. $\Avg{a_{ij}^{\alpha}}$, is given by   Eq. (\ref{av_weight}) that in this case reads 
\bea
\Avg{a_{ij}^{\alpha}}&=&=\frac{e^{-(\omega_{i,\alpha}+\omega_{j,\alpha})+(w_{i,\alpha}+w_{j,\alpha})}}{(e^{w_{i,\alpha}+w_{j,\alpha}}-1)(e^{-(\omega_{i,\alpha}+\omega_{j,\alpha})}+e^{w_{i,\alpha}+w_{j,\alpha}}-1)}.
\label{auncorr}
\eea

Moreover, from Eq. (\ref{probability}) the probability $p_{ij}^{\alpha}$ that the link $(i,j)$ in layer $\alpha$ has weight different from zero is given by 
\bea
p_{ij}^{\alpha}&=&=\frac{\displaystyle{e^{-(\omega_{i,\alpha}+\omega_{j,\alpha})}}}{\displaystyle{e^{-(\omega_{i,\alpha}+\omega_{j,\alpha})}+e^{w_{i,\alpha}+w_{j,\alpha}}-1}}.
\label{puncorr}
\eea
Finally, the probability of a multiplex network $\vec{G}$ in this ensemble, characterized by the $M$ adjacency matrices ${\bf a}^{\alpha}$, is given by  
\bea
P(\vec{G})=\prod_{\alpha=1}^M\prod_{i<j}\pi_{ij}(a_{ij}^{\alpha}),
\eea
with the marginals $\pi_{ij}^{\alpha}(a_{ij}^{\alpha})$ given by  Eq. $(\ref{pib})$

Therefore the entropy ${\cal S}$ of this canonical multiplex ensemble is given by 
 \begin{equation}
\label{entropyS}
\mathcal{S}=-\sum_{\alpha=1}^M\sum_{i<j}\sum_{a_{ij}^{\alpha}=0}^{\infty}\pi_{ij}^{\alpha}(a_{ij}^{\alpha})\log (\pi^{\alpha}_{ij}(a_{ij}^{\alpha})),
\end{equation}
with the marginals $\pi_{ij}^{\alpha}(a_{ij}^{\alpha})$ given by Eqs.~$(\ref{pib})$.

\subsection{Multiplex ensemble with given expected multidegree sequence and given expected multistrength sequence}
Here we consider the example of the correlated weighted multiplex ensemble, in which we fix the expected multidegrees $k_{i}^{\vec{m}}$ of node $i$, for each node $i=1,\ldots,N$, for  each $\vec{m}$. Moreover, in addition to these constraints we impose also a given expected strength $s_{i} ^{\vec{m},\alpha}$ for each node $i=1,2,\ldots, N$ and each multilink $\vec{m}$, in each layer $\alpha$ where $m_{\alpha}=1$. The number of constraints is therefore $K=2^M \cdot N + (2^{M-1})\cdot M \cdot N$. In particular, the constraints we are imposing are  
\bea
\sum_{\vec{G}}F_{i}^{\vec{m}}(\vec{G}) P(\vec{G})&=\sum_{\vec{G}}\left ( \sum_{j\ne i}A_{ij}^{\vec{m}}\right)P(\vec{G})=k_{i}^{\vec{m}}\\
\sum_{\vec{G}}F_{i, \alpha}^{\vec{m}}(\vec{G}) P(\vec{G})&=\sum_{\vec{G}}\left ( \sum_{j\ne i}A_{ij}^{\vec{m}}a_{ij}^{\alpha}\right)P(\vec{G})=s_{i, \alpha}^{\vec{m}},
\label{constraints2}
\eea
where we have now used the multiadjacency matrices $A_{ij}^{\vec{m}}$ with elements given by 
\bea
A_{ij}^{\vec{m}}=\prod_{\alpha=1}^M \left\{\theta(a_{ij}^{\alpha})m_{\alpha}+[1-\theta(a_{ij}^{\alpha})](1-m_{\alpha})\right\}.
\eea

Here we introduce the Lagrangian multipliers $\omega_{i}^{\vec{m}}$ for the first set of constraints and the Lagrangian multipliers $w_{i,\alpha}^{\vec{m}}$ for the second set of constraints. Without loss of generality, if $m_{\alpha}=0$ we set $w_{\alpha}^{\vec{m}}=1/2$. We can do this because the probability of a multiplex network does not depend on any of these values, and we need to define this Lagrangian multipliers only for simplifying the notation.
Using these expression for the Lagrangian multipliers, we obtain the following expression for the probability $P(\vec{G})$ of the multiplex network in the ensembles 
\bea
P(\vec{G})&=&\frac{1}{Z}\exp \left [ -\sum_{\vec{m}}\sum_{i=1}^N \sum_{j \ne i} \left ( \omega_i^{\vec{m}}A_{ij}^{\vec{m}}+
\sum_{\alpha=1}^M w_{i,\alpha}^{\vec{m}} A_{ij}^{\vec{m}} a_{ij}^{\alpha}\right )\right]\nonumber \\
&=&\frac{1}{Z}\exp \left [ -\sum_{i<j}\sum_{\vec{m}}(\omega_i^{\vec{m}}+\omega_j^{\vec{m}})A_{ij}^{\vec{m}}-\sum_{i<j} \sum_{\vec{m}}\sum_{\alpha=1}^M (w_{i,\alpha}^{\vec{m}}+w_{j,\alpha}^{\vec{m}})A_{ij}^{\vec{m}} a_{ij}^{\alpha}\right].
\eea
The partition function $Z$ can be expressed explicitly as
\bea
Z&=&\prod_{i<j}\mathcal{Z}_{ij},
\eea
where $\mathcal{Z}_{ij}$ is given by  
\begin{equation}
\mathcal{Z}_{ij}=\sum_{\vec{m}}e^{-(\omega_i^{\vec{m}}+\omega_j^{\vec{m}})}\prod_{\alpha=1}^{M}\left ( \frac{e^{-(w_{i,\alpha}^{\vec{m}} +w_{j,\alpha}^{\vec{m}}  )}}{1-e^{-(w_{i,\alpha}^{\vec{m}} +w_{j,\alpha}^{\vec{m}}  )}}\right )^{m_{\alpha}}.
\end{equation}
Finally, the Lagrangian multipliers are fixed by the conditions given by Eqs. $(\ref{constraints2})$

We now indicate with $\vec{a}_{ij}$ the vector $(a_{ij}^1,a_{ij}^2,\ldots, a_{ij}^{\alpha},\ldots, a_{ij}^M)$.
The probability of a multiplex network $P(\vec{G})$ can be rewritten as
\begin{equation}
P(\vec{G})=\prod_{i<j}\pi_{ij}(\vec{a}_{ij}),
\label{probability_wmultiplex}
\end{equation}
where the probability of a specific multiweight $\vec{a}_{ij}$ in between nodes  $(i,j)$ is
\begin{equation}
\pi_{ij}(\vec{a}_{ij})=\frac{e^{-(\omega_i^{\vec{m}^{ij}}+\omega_j^{\vec{m}^{ij}})}\prod_{\alpha=1}^{M}\left ( e^{-(w_{i,\alpha}^{\vec{m}^{ij}}+w_{j,\alpha}^{\vec{m}^{ij}})a_{ij}^{\alpha}}\right )^{m_{\alpha}^{ij}}}{\mathcal{Z}_{ij}},
\end{equation}
where $\vec{m}^{ij}=(m^{ij}_1,\ldots, m^{ij}_{\alpha},\ldots, m^{ij}_m)$ with $m^{ij}_{\alpha}=\theta(a_{ij}^{\alpha})$. 
We note here that $\pi_{ij}(\vec{a}_{ij})$ satisfies the following normalization condition:
\begin{equation}
\sum_{\vec{a}_{ij}}\pi_{ij}(\vec{a}_{ij})=1.
\label{piijnorm}
\end{equation}
The average weight $\Avg{a_{ij}^{\alpha}A_{ij}^{\vec{m}}}$ of multilink $\vec{m}$ between nodes $(i$ and $j$ in layer $\alpha$ and the probability $p_{ij}^{\vec{m}}$ of multilink $\vec{m}$ between nodes $i$ and $j$ are given by 
\bea
\Avg{a_{ij}^{\alpha}A_{ij}^{\vec{m}}}&=&\frac{e^{-(\omega_{i}^{\vec{m}}+\omega_{j}^{\vec{m}})}}{\mathcal{Z}_{ij}} \left ( \frac{1}{1-e^{-(w_{i,\alpha}^{\vec{m}}+w_{j,\alpha}^{\vec{m}})}}\right )\prod_{\beta=1}^{M}\left ( \frac{e^{-(w_{i,\beta}^{\vec{m}}+w_{j,\beta}^{\vec{m}})}}{1-e^{-(w_{i,\beta}^{\vec{m}}+w_{j,\beta}^{\vec{m}})}}\right )^{m_{\beta}},\\
p_{ij}^{\vec{m}}&=&\frac{e^{-(\omega_i^{\vec{m}}+\omega_j^{\vec{m}})}}{\mathcal{Z}_{ij}}\prod_{\alpha=1}^{M}\left ( \frac{e^{-(w_{i,\alpha}^{\vec{m}} +w_{j,\alpha}^{\vec{m}})}}{1-e^{-(w_{i,\alpha}^{\vec{m}} +w_{j,\alpha}^{\vec{m}}  )}}\right )^{m_{\alpha}}.
\eea

Finally, since the probability of a multiplex network $P(\vec{G})$ is given by Eq. (\ref{probability_wmultiplex}), the entropy ${\cal S}$ defined in Eq. $(\ref{entropy})$ in this ensemble is given by 
\bea
{\cal S}&=-\sum_{i<j}\sum_{\vec{a}_{ij}} \pi_{ij}(\vec{a}_{ij})\log  \pi_{ij}(\vec{a}_{ij}).
\label{entropy2}
\eea

\section{Background information on Figure 4 of the main text}
\label{four}
As an example of a possible application of the indicators $\Psi$ and $\Xi$, we analyzed a case inspired by the CoCi-PRE multiplex network. Due to the numerical limitations of the programs that are able to evaluate the entropy of multiplex ensembles, we performed a finite-size analysis of the indicators $\Psi$ and $\Xi$ as a function of the size of the multiplex network $N=128,256,\ldots, 2048$.
In particular, we considered the following  undirected multiplex ensembles:    
\begin{itemize}
\item {\it  Correlated weighted multiplex ensemble.}
First, we created the correlated multiplex ensemble with power-law multidegree distributions with exponents $\gamma^{(1,m_2)}=2.6$ for $m_2=0,1$ and $\gamma^{(0,1)}=1.9$ (for multidegree $(0,1)$ we imposed a structural cut-off).  
In particular, in order to avoid the effects of fluctuations in the multidegree sequence, we ranked the multidegrees as $r=1,2,\ldots N$ and taken the degree sequence in which the multidegree $k^{\vec{m}}_r$ of rank $r$ is defined by 
\bea
\frac{r}{N}=\int_{k_r^{\vec{m}}}^{K}P(k^{\vec{m}})dk^{\vec{m}},
\eea
where we have taken the maximal cut-off $K=N$ for $\gamma^{\vec{m}}>2$ and $K=\sqrt{\Avg{k^{\vec{m}}}N}$ for $\gamma^{\vec{m}}<2$.
Moreover, multistrengths were assumed to satisfy
\bea
s_i^{\vec{m},\alpha}&=&c_{\vec{m},1}(k^{\vec{m},\alpha})^{\lambda_{\vec{m},\alpha}}
\label{skm}
\eea
with 
$c_{\vec{m},\alpha}=1$ and $\beta^{(1,m_2),1}=1,$ for $m_2=0,1$; $\beta^{(1,1), 2}=1.3$, $\beta^{(0,1),2}=1.1$.
\item {\it  Uncorrelated weighted multiplex ensemble.}
In this ensemble, we set the expected degree $k_i^{\alpha}$ of every node $i$ in every layer $\alpha=1,2$ to be equal to the sum of the multidegrees (with $m_{\alpha}=1$) in the correlated weighted multiplex ensemble. Moreover, we set the expected strengths $s_i^{\alpha}$ of every node $i$ in every layer $\alpha$ to be equal to the sum of the multistrengths of node $i$ in layer $\alpha$ in the correlated weighted multiplex ensemble. 
\end{itemize}
\begin{figure}
\begin{center}
\centerline{\includegraphics[width=8in]{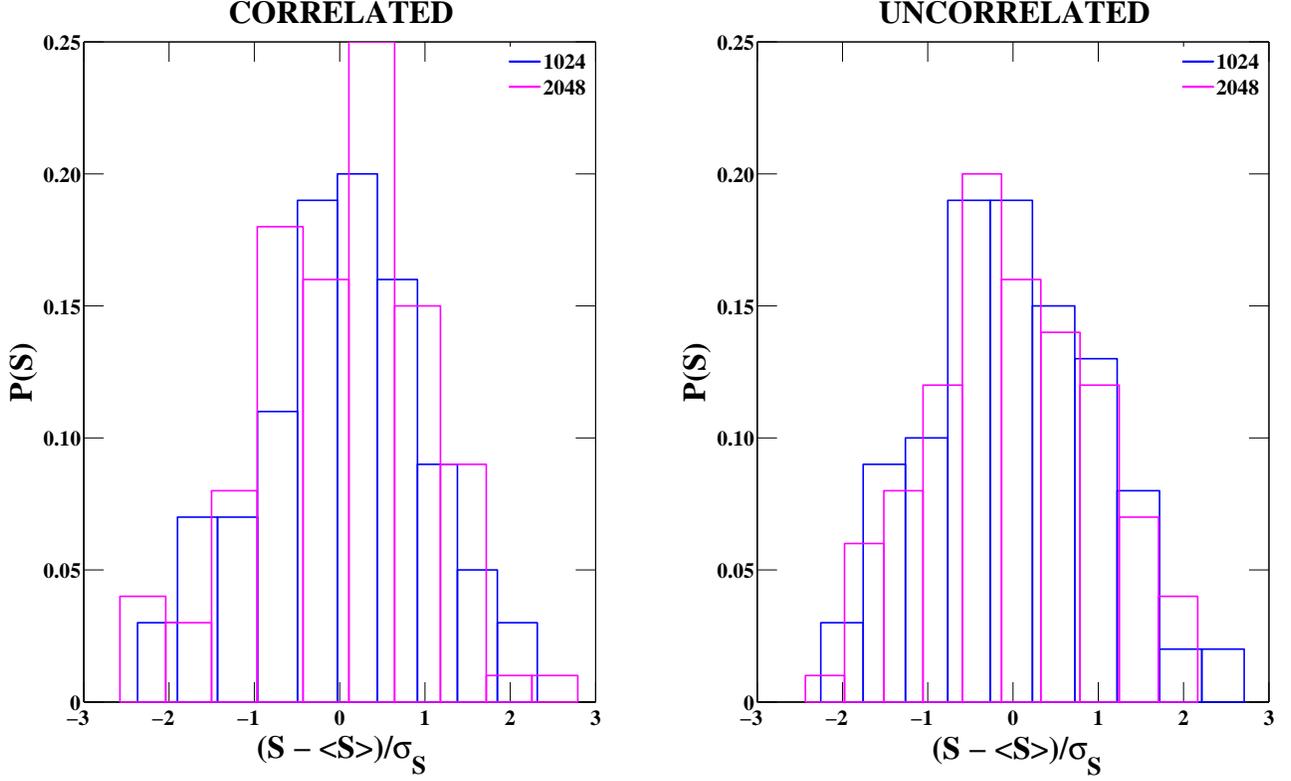}    }
\end{center}
\caption{The $P({\cal S})$ distribution in the null models for correlated and uncorrelated multiplex ensembles in which the weights are distributed uniformly over the links of the multiplex network. The $P(S)$ distributions are calculated over $100$ randomizations of the weights for multiplex networks of $N=1024$ and $N=2048$ nodes.}
\label{figps}
\end{figure}
\begin{figure}
\begin{center}
\centerline{\includegraphics[width=8in]{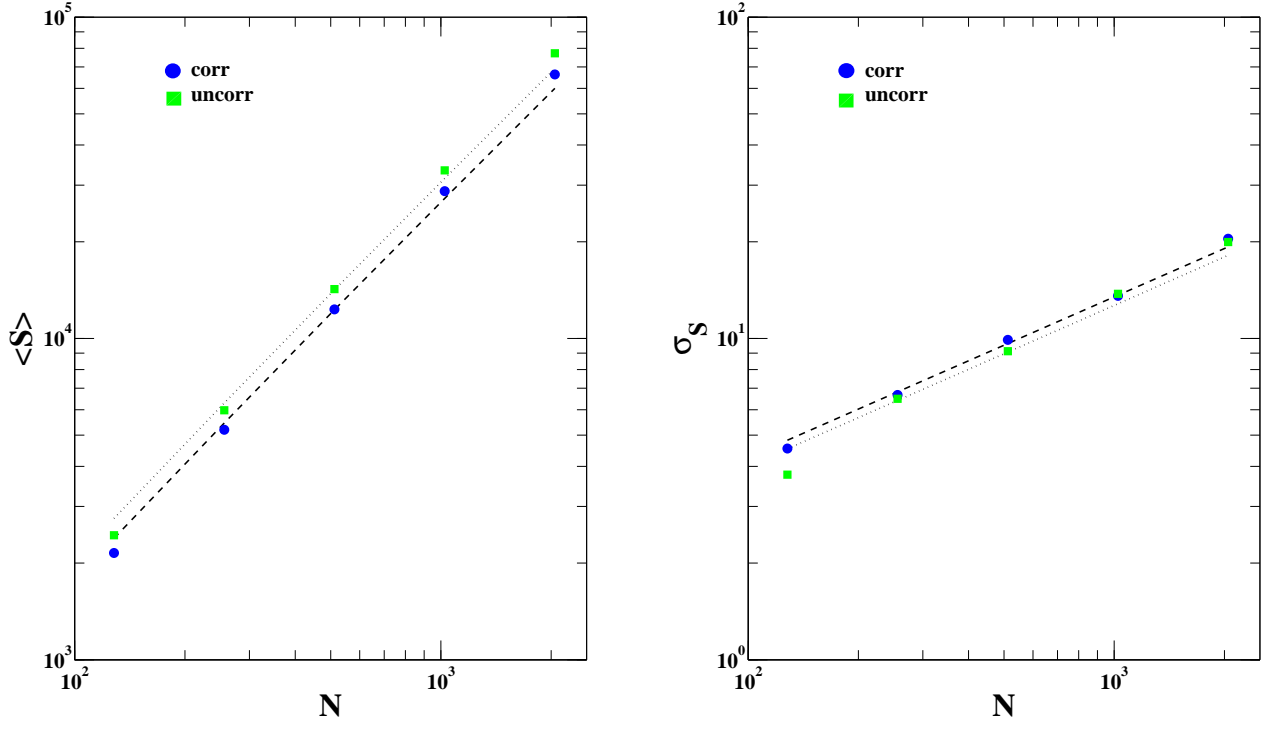} }
\end{center}
\caption{The mean $\Avg{\cal S}$ and variance $\sigma_S$ as a function of the system size $N$ for the null models of correlated and uncorrelated multiplex ensembles in which the weights are distributed uniformly over the links of the multiplex network. The solid lines indicate the fit of the data in which we assume $\Avg{{\cal S}}=a N\log N$ and $\sigma_S=b\sqrt{N}$.}
\label{figps2}
\end{figure}
We measured the indicator $\Psi$ that compares the entropy of a weighted multiplex ensemble $\cal S$ with the entropy of a weighted multiplex ensemble in which weights are distributed homogeneously. Therefore, $\Psi$ can be defined as   
\bea
\Psi=\frac{|{\cal S}-\avg{\cal S}_{\pi(w)}|}{\Avg{(\delta {\cal S})^2}_{\pi(w)}},
\eea
where the average $\avg{\ldots}_{\pi(w)}$ is calculated over multiplex networks with the same structural properties but with weights distributed homogeneously. In particular, when the weight distribution is randomized, the multiplex networks are constrained in such a way that each link must have a minimal weight (i.e. $w_{ij}>1$), while the remaining of the total weight is distributed randomly across links.
When numerically evaluating $\Avg{\ldots}_{\pi(w)}$, we obtained the average over $100$ weight randomizations. 
The distribution $P({\cal S})$ of the entropy ${\cal S}$ calculated over these randomizations, both for the uncorrelated weighted multiplex ensemble and for the correlated weighted multiplex ensemble, is shown in Figure $\ref{figps}$. In both cases, we observe a distribution that can be fitted by a Gaussian function with mean and variance scaling as $\Avg{{\cal S}}\propto N\log N$ and $\Avg{(\delta {\cal S})^2}_{\pi(w)}\propto \sqrt{N}$ (See Figure $\ref{figps2}$).
We call $\Psi^{corr}$ the indicator $\Psi$ calculated on the correlated multiplex ensemble and indicate with $\Psi^{corr}$ the indicator $\Psi$ calculated on the corresponding uncorrelated multiplex ensemble. Finally, to quantify the additional amount of information carried by the correlated multiplex ensemble with respect to the uncorrelated multiplex ensemble, we measured the indicator $\Xi$ as
\bea
\Xi=\frac{\Psi^{corr}}{\Psi^{uncorr}}.
\eea 
The finite-size scaling of $\Psi^{corr},\Psi^{corr}$ and $\Xi$ are shown in Figure 4 in the manuscript.

\end{document}